\documentclass[useAMS,usenatbib]{mnras}
\title[Kinematics and populations of early-type galaxies]{Integral-field kinematics and stellar populations of early-type galaxies out to three half-light radii}
\author[N. F. Boardman et al.]{Nicholas Boardman$^{1}$\thanks{E-mail: nfb@st-andrews.ac.uk},
Anne-Marie Weijmans$^{1}$, Remco van den Bosch$^{2}$, \and Harald Kuntschner$^{3}$, Eric Emsellem$^{3,4}$,
Michele Cappellari$^{5}$, Tim de Zeeuw$^{3,6}$, \and Jesus Falc{\'o}n-Barroso$^{7,8}$,
Davor Krajnovi\'c$^{9}$, Richard McDermid$^{10,11}$, Thorsten Naab$^{12}$, \and Glenn van de Ven$^{2}$, 
Akin Yildirim$^{2}$\\
$^{1}$School of Physics and Astronomy, University of St Andrews, KY16 9SS UK\\
$^{2}$Max Planck Institute for Astronomy, K\"{o}nigstuhl 17, D-69117 Heidelberg, Germany\\
$^{3}$European Southern Observatory, Karl-Schwarzschild-Str. 2, 85748 Garching, Germany\\
$^{4}$Universit\'e Lyon 1, Observatoire de Lyon, Centre de Recherche Astrophysique de Lyon \\ \hspace*{0.5cm} and Ecole Normale Sup\'erieure de Lyon, 9 avenue Charles Andr\'e, F-69230 Saint-Genis Laval, France\\
$^{5}$Sub-department of Astrophysics, Department of Physics, University of Oxford, Denys Wilkinson Building, Keble Road, Oxford OX1 3RH\\
$^{6}$Sterrewacht Leiden, Leiden University, Postbus 9513, 2300 RA, Leiden, The Netherlands\\
$^{7}$Instituto de Astrofisica de Canarias, E-38200, La Laguna, Spain\\
$^{8}$Depto. Astrofisica, Universidad de La Laguna (ULL), E-38206 La Laguna, Tenerife, Spain\\
$^{9}$Leibniz-Institut f\"ur Astrophysik Potsdam (AIP), An der Sternwarte 16, D-14482 Potsdam, Germany \\
$^{10}$Department of Physics and Astronomy, Macquarie University, Sydney, NSW 2109, Australia\\
$^{11}$Australian Astronomical Observatory, PO Box 915, North Ryde, NSW 1670, Australia\\
$^{12}$Max-Planck-Institut f\"ur Astrophysik,
Karl-Schwarzschild-Str. 1, 85741 Garching, Germany\\}

\newcommand{\HI}{{\sc H\,i}}
\newcommand{\OIII}{{\sc O\,iii}}
\newcommand{\NI}{{\sc N\,i}}
\newcommand{\lagr}{\mathcal{L}}
\usepackage{anysize}
\usepackage{graphicx}
\usepackage{amsmath}
\usepackage[toc,page]{appendix}
\usepackage[normalem]{ulem}
\usepackage[]{hyperref}
\marginsize{2cm}{2cm}{2cm}{2cm}
\begin{document}

\date{Accepted ???. Received ???; in original form ???}

\pagerange{\pageref{firstpage}--\pageref{lastpage}} \pubyear{2016}

\maketitle

\label{firstpage}

\begin{abstract}
We observed twelve nearby \HI -detected  early-type galaxies (ETGs) of stellar mass $\sim 10^{10}M\odot \leq  M_* \leq \sim 10^{11}M\odot$ with the Mitchell Integral-Field Spectrograph, reaching approximately three half-light radii in most cases. We extracted line-of-sight velocity distributions for the stellar and gaseous components. We find little evidence of  transitions in the stellar kinematics of the galaxies in our sample beyond the central effective radius, with centrally fast-rotating galaxies remaining fast-rotating and centrally slow-rotating galaxies likewise remaining slow-rotating. This is consistent with these galaxies having not experienced late dry major mergers; however, several of our objects have ionised gas that is misaligned with respect to their stars, suggesting some kind of past interaction. We extract Lick index measurements of the commonly-used H$\beta$, Fe5015, Mg\,{\it b}, Fe5270 and Fe5335 absorption features, and we find most galaxies to have flat H$\beta$ gradients and negative Mg\,{\it b} gradients. We measure gradients of age, metallicity and abundance ratio for our galaxies using spectral fitting, and for the majority of our galaxies find negative age and metallicity gradients. We also find the stellar mass-to-light ratios to decrease with radius for most of the galaxies in our sample. Our results are consistent with a view in which intermediate-mass ETGs experience mostly quiet evolutionary histories, but in which many have experienced some kind of gaseous interaction in recent times.

\end{abstract}

\begin{keywords}
galaxies: elliptical and lenticular, cD - galaxies: ISM - galaxies: kinematics and dynamics - galaxies: structure - galaxies: evolution - ISM: kinematics and dynamics
\end{keywords}

\section{Introduction}\label{intro}

The evolutionary paths of lenticular (S0) and elliptical (E) galaxies, collectively referred to as early-type galaxies (ETGs), continue to be of great interest, with ETGs commonly thought to represent the endpoints of galaxy evolution. ETG imaging has repeatedly shown a dramatic size evolution since redshift $z \simeq 2$, with massive ETGs (those with stellar masses $M_* \geq \sim 10^{11}M\odot$)  significantly smaller and more compact in the past than in the present day \citep{trujillo2006,cimatti2012,vandokkum2010, vandokkum2013}. This size evolution can be explained by a ``two-phase" picture of ETG formation \citep{oser2010,oser2012}. In this picture, a dense core is formed at early times ($z \simeq 2$) via strongly dissipative processes. Star formation is quenched subsequently, and then a stellar halo is built up over time from dry merging and accretion episodes. 

Further information on ETGs may be obtained from spectroscopic observations. In particular, two-dimensional spectroscopy from the SAURON instrument \citep{bacon2001} by the SAURON \citep{dezeeuw2002} and ATLAS\textsuperscript{3D} \citep{cappellari2011} surveys have greatly enhanced our understanding of ETGs' structures. ETGs can be slow-rotating or fast-rotating, with a wide range of kinematic substructures reported in individual systems \citep{kraj2011}; as such, ETGs may be classed as``fast rotators" (FRs) or ``slow rotators" (SRs), based on their line-of-sight kinematics \citep{emsellem2007,emsellem2011}. More massive ETGs are also more likely to be SRs, while less massive ones are more likely to be FRs \citep{emsellem2011}. 

A key finding is that FRs, which dominate the ETGs population below a characteristic galaxy stellar mass $ M_{crit} \simeq 2\times 10^{11}M_\odot$, have a velocity dispersion $\sigma$ that correlates well with galaxy properties like colour, age, the stellar initial mass function (IMF) and the molecular gas fraction \citep{cappellari2013a}. Given that $\sigma$ correlates with the galaxy bulge mass fraction, this suggests that the evolution of fast-rotating ETGs is linked closely to central mass-growth; slow-rotating ETGs, which dominate at masses above $M_{crit}$, appear to evolve in some other way. The structure of fast-rotator ETGs was also shown to parallel  that of spiral galaxies \citep{cappellari2011a} and to lie near-parallel with respect to spirals on the mass-size plane, with ETGs systematically smaller (i.e. more concentrated) at a given mass \citep{cappellari2013}. These findings are consistent with a view in which ETGs form via two main channels: fast rotators start as star forming disks and grow their bulges via dissipative processes, followed by quenching, while the more massive slow rotators form as in the two-phase scenario described above, with an early rapid dissipative formation followed by repeated dry merger events \citep[see][for a review.]{cappellari2016}

The exact nature of the ETG-spiral link remains unclear. Results from the CALIFA IFU survey have shown that ETGs on average have less angular momentum and are more centrally concentrated than spiral galaxies \citep{falconbarroso2015}, which is consistent with results from major merger simulations \citep{querejeta2015}. \citet{querejeta2015} point out that some spiral galaxies (particularly Sa galaxies) do indeed overlap with ETGs on the momentum-concentration plane, which is consistent with passive fading scenarios \citep[e.g.][]{dressler1997,peng2015}. Such scenarios are also consistent with measurements of total luminosity and disk scale length, which are larger for Sa galaxies than S0s \citep{vaghmare2015}. At the same time, around half of local ETGs display significant star-gas misalignments both for neutral gas \citep{serra2014} and also for ionised gas \citep[e.g.][]{sarzi2006,davis2011}; this implies that some kind of interaction event must have occurred in a large fraction of ETGs. 

One way to study ETGs further is to measure their stellar kinematics to $2R_e$ and beyond - something not possible for much of the ATLAS\textsuperscript{3D} and CALIFA datasets - in order to test if the kinematics are consistent with currently proposed formation scenarios.  \citet{arnold2014} present slitlet stellar kinematics for 22 ETGs from the SLUGGS survey \citep{brodie2014}  out to $~2-4 R_e$ and find kinematic transitions in a few of their objects, with abrupt drops in the angular momentum beyond  $1R_e$; \citet{arnold2014} argue this to be evidence of two-phase formation in these objects, with the momentum drops signifying a transition to accretion-dominated stellar halos. By contrast, \citet{raskutti2014}, do not report finding such transitions from Mitchell Spectrograph IFU observations of their massive ETGs out to $~2-5R_e$. In addition, galaxies produced in cosmological simulations rarely display such drops in momentum, even for galaxies with significant dry accretion at late times \citep{naab2014,wu2014}. 

Further information on ETGs' histories is encoded within the properties of their stellar populations. Most ETGs' stellar light is dominated by old stars, as expected from their ``red and dead" appearance. In addition, compact ETGs are older, more metal-rich and more alpha-enhanced than more extended ETGs of the same mass; this can be understood as being a result of dissipative galaxy mergers \citep{cappellari2013a,mcdermid2015}, which drive central galaxy formation and so increase the compactness of a galaxy \citep{khochfar2006}. No dependence on size is found for ETGs' stellar population properties at a given velocity dispersion $\sigma$, suggesting that $\sigma$ (and not mass) is the strongest individual indicator of ETGs' stellar populations \citep{cappellari2006,graves2009,cappellari2013a}. At a given dispersion $\sigma$, galaxies with a more massive black hole (BH) appear to be older and more alpha-enhanced \citep{navarro2016}, suggesting that the BH of a galaxy must also play a role; for massive active galaxies in particular, the central BH likely plays a significant part in their host galaxies' quenching \citep[see][for a review.]{fabian2012}

Within individual ETGs, various radial gradients have been reported for their stellar population properties in recent years. ETGs have repeatedly been found to have negative metallicity gradients \citep[e.g.][]{davies1993,rawle2008,greene2013,greene2015,wilkinson2015,scott2013}, along with age gradients that are flat or close to flat \citep[e.g.][]{rawle2008,wilkinson2015,scott2013}. The metallicity gradients of some ETGs have also been found to change with radius, reflecting their different formation histories. Massive ETGs have been consistently reported to show flattening metallicity gradients beyond $\simeq 1R_e$, consistent with the gradients being ``washed out" by repeated dry accretion events \citep[e.g.][]{coccato2010, greene2013,pastorello2014}, while less massive ETGs show no such feature \citep{weijmans2009,pastorello2014}. However, only a handful of galaxies significantly below $M_{crit}$ have been studied in this way to date, making it unclear if this picture holds for \textit{all} low-mass ETGs; if low-mass ETGs are linked to spiral galaxies and have evolved via mainly internal processes, then this should indeed be the case.

It is also interesting to consider how the stellar mass-to-light ratio $M_*/L$ varies across a given ETG. $M_*/L$ is typically assumed constant across ETGs for dynamical modelling purposes. However, the above-described stellar population gradients imply that gradients in $M_*/L$ should also be expected. \citet{tortora2011} fit stellar population models to Sloan Digital Sky Survey \citep[SDSS;][]{ahn2014} imaging out to 1$R_e$, and find shallow $M_*/L$ drops for ETGs with old centers and shallow rises for ETGs with young centers. Stellar population modelling of spectra out to multiple $R_e$ is a powerful way in which to investigate this point further.

The Mitchell Integral-Field Spectrograph \citep{hill2008}, on the Harlan J. Smith telescope, is an ideal tool for investigating ETGs' structures beyond 1 $R_e$.  It has a $1.68 \times 1.68$ arcminute field of view, and uses large individual optical fibres (radius 2.08$''$) which are well-optimised for deep observations of objects and which reduce the need for spatial binning. 

Here, we present results for twelve nearby $\sim 10^{10}M\odot \leq  M_* < M_{crit}$  ETGs observed with the Mitchell Spectrograph. Our observations reach past 2$R_e$ in most cases. We present stellar and gaseous kinematics and absorption line index measurements. We also present results from stellar population modelling.  All of our ETGs have previously been observed as part of the ATLAS\textsuperscript{3D} survey,  making our own observations greatly complementary.

The structure of this paper is as follows. We discuss our observations and data reduction in \autoref{sample}. We describe the extraction of stellar and ionised gas kinematics in \autoref{kin}, wherein we also parametrise the stellar angular momentum, of our galaxies as a function of radius. We describe the extraction of absorption line indices in  \autoref{linestrength}, and we then explain the calculation of stellar population parameters through spectral fitting in \autoref{popsppxf}. We discuss our findings in \autoref{disc} and we conclude in \autoref{sum}.

\section{Sample and data reduction}\label{sample}

\begin{table*}
\begin{center}
\begin{tabular}{|c|c|c|c|c|c|c|c|c|c|c|}
\hline 
Galaxy & RA & DEC & T-type & FR/SR & $R_e ('')$ & $M_K$ & Distance (Mpc) & $\log_{10}(M_*)$ \\ 
\hline 
NGC 680 & 27.447035 &  21.970827 & -4.0 & FR & 14.5 & -24.17 & 37.5 & 11.09 \\ 
\hline 
NGC 1023 & 40.100052 & 39.063251 & -2.7 & FR & 47.9 & -24.01 & 11.1 & 11.02 \\ 
\hline 
NGC 2685 & 133.894791 & 58.734409  &  -1.0 & FR & 25.7 & -22.78 & 16.7  & 10.48 \\ 
\hline 
NGC 2764 & 137.072983 & 21.443447 &  -2.0 & FR & 12.3 & -23.19 & 39.6 & 10.66 \\ 
\hline 
NGC 3522 & 166.668549 & 20.085621 &  -4.9 & SR & 10.2 & -21.67 & 25.5 & 9.99 \\ 
\hline 
NGC 3626 & 170.015808 & 18.356791 &  -1.0 & FR & 25.7 & -23.30 & 19.5 & 10.71 \\ 
\hline 
NGC 3998 & 179.484039 & 55.453564 &  -2.1 & FR & 20.0 & -23.33 & 13.7 & 10.73 \\ 
\hline 
NGC 4203 & 183.770935 & 33.197243 &  -2.7 & FR  & 29.5 & -23.44 & 14.7 & 10.77 \\ 
\hline 
NGC 5582 & 215.179703 & 39.693584 &  -4.9 & FR & 27.5 & -23.28 & 27.7 & 10.70 \\ 
\hline 
NGC 5631 & 216.638687  & 56.582664 &  -1.9 & SR & 20.9 & -23.70 & 27.0 & 10.89 \\ 
\hline 
NGC 6798 & 291.013306 & 53.624752 &  -2.0 & FR & 16.9 & -23.52 & 37.5 & 10.80 \\ 
\hline 
UGC 03960 & 115.094856 & 23.275089 &  -4.9 & SR & 17.4 & -21.89 & 33.2 & 10.09 \\ 
\hline
\end{tabular} 
\end{center}
\caption{Summary of our Mitchell Spectrograph sample of ETGs. RA, DEC, $R_e$, T-type, K-band magnitude $M_K$ and distances are from \citet{cappellari2011} and references therein. $\log_{10}(M_*)$ was calculated using the K-band magnitudes, after applying Equation 2 in \citet{cappellari2013b}. The FR/SR classifications are the 1$R_e$ classifications reported in \citet{emsellem2011}.}
\label{tab1}
\end{table*}

\begin{figure}
\begin{center}
	\includegraphics[trim = 2cm 2cm 7cm 17cm,scale=0.6]{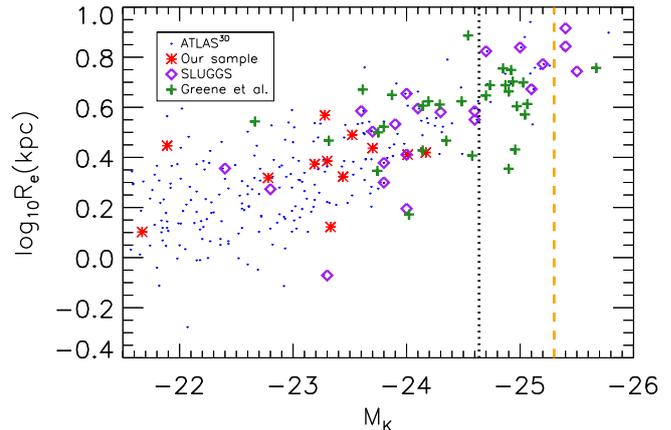}
	\caption{Plot of $R_e$ against absolute K-band magnitude for the full ATLAS\textsuperscript{3D} ETG sample (blue dots), our selected ETGs overlaid (red crosses), the SLUGGS sample (purple diamonds) and the sample of \citet{raskutti2014} (green pluses). The vertical dotted line corresponds to $M_{crit} = 2\times 10^{11} M_{\odot}$, using Equation 2 from \citet{cappellari2013b}, while the orange dashed line corresponds to the upper magnitude limit of the MASSIVE survey \citep{ma2014}. Our sample focusses on a somewhat lower mass region than SLUGGS, \citet{greene2013} or MASSIVE.}
	\label{sample1}
	\end{center}
\end{figure}

In \autoref{tab1}, we present our sample of 12 nearby ETGs. We manually selected our sample from the ATLAS\textsuperscript{3D} survey on the basis of detected \HI\ emission from the Westerbork Synthesis Radio Telescope (WSRT) \citep{serra2012}. NGC 680 and NGC 1023 are both classified as ``u" in \citet{serra2012}, indicating substantially unsettled \HI\ morphology; the remaining 10 sample galaxies contain large-scale regularly-rotating \HI\ structures (the ``D" galaxies in \citet{serra2012}). \citet{serra2014} present \HI\ velocity maps for all \HI\ detected ATLAS\textsuperscript{3D} galaxies, including the galaxies in our sample, so we direct the interested reader to that paper.

In \autoref{sample1}, we compare our sample to those of ATLAS\textsuperscript{3D} and SLUGGS \citep{arnold2014}, as well as with the massive ETG sample presented in \citet{greene2013} and \citet{raskutti2014}. {We also show the upper magnitude limit of the MASSIVE ETG survey \citep{ma2014} on the same figure. We show that the masses of our sample tend towards lower values than the latter {three samples}, making our sample different from other wide-field samples published thus far.  Our sample galaxies all have masses which fall below or around $10^{11} M_\odot$. We obtained $M_K$ and $R_e$ values for our sample from Table 3 of \citet{cappellari2011}, converting the latter to units of kpc using the distances provided in that table; we obtained $R_e$ values for the full ATLAS\textsuperscript{3D} sample in the same manner. We obtained $R_e$ values, distances and $M_K$ quantities for the SLUGGS survey from Table 1 of \citet{arnold2014}. We extracted $R_e$ and $M_K$ quantities for the \citet{raskutti2014} objects directly from their Figure 1.

Observations were taken on the Mitchell Spectrograph using the VP2 grating over 27 nights, spaced over four observing runs. Galaxies were observed using a series of thirty-minute exposures, with bracketing sky observations of 15 minutes also taken to enable sky subtraction. Bias frames, flat frames and arc frames were taken at the beginning and end of each night, with Ne+Cd comparison lamps used for the arc frames in all cases. We observed some galaxies longer than others, owing to observing time constraints. We also chose to observe certain galaxies over multiple pointings, to better-capture their structure beyond the central effective radius. 

The Mitchell spectrograph has a one-third fill factor, and complete sky coverage can be achieved by taking observations over three dither positions. We observed most galaxies over all three dither positions in order to obtain full coverage. However, a couple of galaxies could not be observed in this way due to observing time constraints: NGC 4203 was observed over two dither positions instead, while NGC 2685 was also observed over two dither positions over one of its two pointings.

We reduced all data using the custom-built VACCINE pipeline \citep{adams2011}. VACCINE subtracts science frames of overscan and bias, traces fibres along the CCD chip, calculates wavelength solutions, constructs sky frames using bracketing sky exposures and detects cosmic rays, before collapsing the output science images into a set of 1D spectra. It uses techniques similar to those proposed by \citet{kelson2003} in order to avoid the resampling of data and, in turn, correlated errors. 

We perform sky-subtraction at this point by averaging over bracketing sky observations, for the purpose of quality control and to estimate the signal-to-noise (S/N) of galaxy spectra. This approach implicitly assumes that the sky varies linearly as a function of time; however, sky variations will actually be non-linear in practice, which can become significant in cases where the sky spectrum dominates the observed light \citep[e.g.][]{blanc2013}. Therefore, we will later fit the sky spectra as part of the kinematic extraction discussed in \autoref{kin}.

We found that some fibres fell off the CCD chip, and also found that certain fibres had insufficient wavelength ranges for the bright 5400.56\AA\ Ne line to be visible; we masked such fibres in our analysis, to enable a consistent wavelength range of 4800-5400\AA\ and to make the derived wavelength solutions as robust as possible. All such fibres were located near the edge of the CCD chip, so removing them has a minimal effect on our analysis.

We combined all science frames into a single spectral datacube for each galaxy, with all spectra for a given galaxy interpolated onto a common linear wavelength scale. We masked out fibre positions with excessively low S/N in all cases. We calculated the instrumental resolutions of the galaxy spectra by fitting Gaussians to the 5154.660 and 5400.56 \AA\ emission lines in the master arc frames, weighting the frames  in accordance with that galaxy's observation dates. We found the spectral resolution to vary smoothly as a function of fibre position; we therefore chose to fit a fifth-order polynomial to the resolution as a function of fibre position to eliminate the noise inherent in this calculation. The lowest spectral full-width at half-maximum (FWHM) that we found from this process is 1.2\AA\, with the maximum derived value being 1.7\AA\; this latter value corresponds to an instrumental velocity dispersion of approximately $42$ km/s.

In \autoref{sample2} we compare the radial extent of our data, parametrised via the $R_{\rm max}$ parameter - the maximum aperture radius of the data \citep[e.g.][]{emsellem2011}  - with that of the ATLAS\textsuperscript{3D} and SLUGGS samples. We define $R_{\rm max}$ to be the maximum radius of a circular aperture with area equal to an ellipse that is at least 85\% filled with spectra; we symmetrised the fibre positions for NGC 1023 and NGC 3626 when calculating $R_{\rm max}$, due to these galaxies not being centred on our FOV. We find our FOV to be comparable with that of SLUGGS for less massive ETGs, with our data extending beyond $2 R_e$ in most cases; our data is therefore highly complementary with respect to the ATLAS\textsuperscript{3D} data. We obtain ATLAS\textsuperscript{3d} $R_{\rm max}$ values from Table B1 in \citet{emsellem2011}, using the same $R_e$ values as before, and we obtain SLUGGS $R_{\rm max}$ values using Table 1 of \citet{arnold2014}.

\begin{figure}
\begin{center}
	\includegraphics[trim = 9cm 12cm 10cm 5cm,scale=0.5]{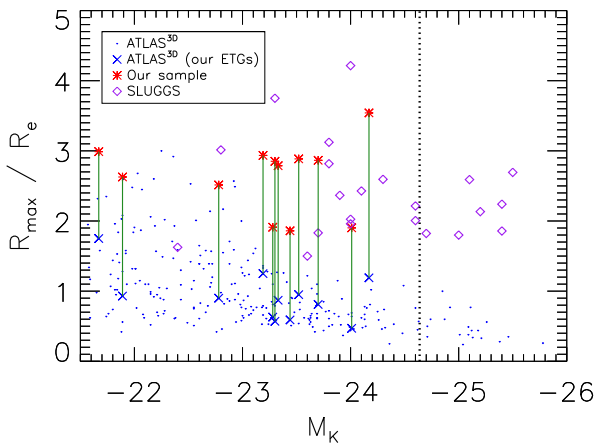}
	\caption{Plot of $R_{\rm max}$/$R_e$. Blue crosses represent the ATLAS\textsuperscript{3D} values for our galaxies, and green lines show the difference between our values and the ATLAS\textsuperscript{3D} ones. All other lines and symbols are as before. Our coverage beyond $2R_e$ in many cases and is comparable to the coverage of SLUGGS over lower-mass ETGs, whereas ATLAS\textsuperscript{3D}  generally reaches approximately $1R_e$.}
	\label{sample2}
	\end{center}
\end{figure}

\begin{table*}
\begin{center}
\begin{tabular}{|c|c|c|c|c|c|}
\hline 
Galaxy & Obs. time (s) & Obs. date &  Pointings & $R_{\rm max}/R_e$ & $R_{\rm max}$ (kpc)\\ 
\hline 
NGC 680 & 9000 & Jan 2011 & 1 & 3.5 &  9.3\\ 
\hline 
NGC 1023 & 39600 & Jan 2011 & 2 & 1.9 & 4.9 \\ 
\hline 
NGC 2685 & 44100 & Jan 2011 & 2 & 2.5 & 5.2\\ 
\hline 
NGC 2764 & 81000 & Mar 2010 & 1 & 2.9 & 6.9\\ 
\hline 
NGC 3522 & 12600 & Jan 2011 & 1 & 3.0 & 3.8\\ 
\hline 
NGC 3626 & 21600 & Apr 2011 & 1 & 2.9 & 6.9\\ 
\hline 
NGC 3998 & 66600 & Mar 2010 & 1 & 2.8 & 3.7 \\ 
\hline 
NGC 4203 & 12600 & Jun 2010 & 1 & 1.9 & 3.9 \\ 
\hline 
NGC 5582 & 59400 & Jun 2010 & 1 & 1.9 & 7.1 \\ 
\hline 
NGC 5631 & 25200 & Apr 2011 &  2 & 2.9 & 7.8 \\ 
\hline 
NGC 6798 & 21800 & Jun 2010 & 1 & 2.9 & 8.9 \\ 
\hline 
UGC 03960 & 18000 & Jan 2011 & 1 & 2.6 & 7.3 \\ 
\hline
\end{tabular} 
\end{center}
\caption{Summary of Mitchell spectrograph data for our ETGs, in terms of observation time, date of observing run, number of pointings and achieved (circular) aperture radius $R_{\rm max}$. The kpc values of $R_{\rm max}$ were calculated using the distances given in \citet{cappellari2011} and references therein.}
\label{tab2}
\end{table*}

We found it necessary to bin spectra in galaxies' outer regions in order to improve the signal-to-noise (S/N) ratios. For each galaxy, we broadened all input spectra to the largest measured FWHM for that galaxy and we binned the spectra using the publicly available Voronoi Binning algorithm \citep{cappellari2003}. We used a target S/N of 20 per spectral pixel ($\simeq 30$ per angstrom) for all galaxies. We detail the observation times, observation date, number of pointings and achieved $R_{\rm max}$ for each galaxy in \autoref{tab2}

\subsection{Flux calibration}

Due to a lack of suitable flux star observations, we were unable to flux-calibrate our Mitchell spectra in the usual way. We therefore performed flux-calibration by comparing our spectra for NGC 3522 to the SDSS spectrum for the same system, and then further comparing our galaxy spectra with those of the ATLAS\textsuperscript{3D} survey. We describe this process in the remainder of this subsection. NGC 3522 is one of two galaxies that we have in common with the Sloan spectral sample, the other being UGC 03960, while all of our galaxies have ATLAS\textsuperscript{3D} spectra.

As we would be using SDSS data to construct our final flux correction curve, we first verified that the SAURON and SDSS datasets were consistent. We took the SDSS spectrum for NGC 3522 and obtained an equivalent SAURON spectrum by summing over a three-arcsecond aperture in order to match the SDSS fibre radius. We matched the spectral resolutions and then smoothed both spectra. We divided the spectra through and fitted a seventh-order polynomial to derive a correction curve.  We find the resulting curve to be almost flat over most of the SAURON wavelength region, as shown in \autoref{sdss}. From this, we conclude that the flux calibration for NGC 3522 is consistent between ATLAS\textsuperscript{3D} and SDSS.

\begin{figure}
\begin{center}
	\includegraphics[trim = 1cm 13cm 0cm 6cm, scale=0.7]{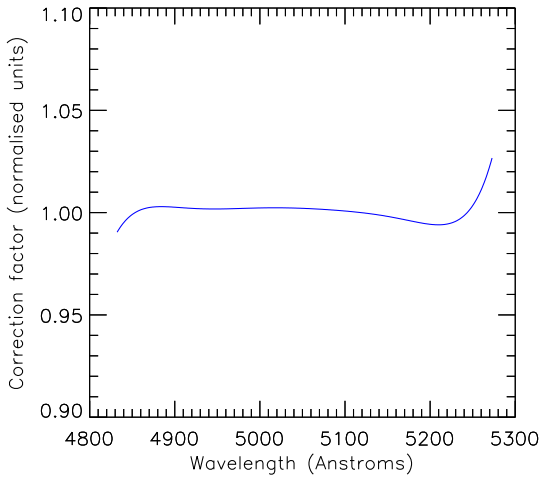}
	\caption{Polynomial SAURON-to-SDSS correction curve obtained for NGC 3522 over the wavelength range of the reduced ATLAS\textsuperscript{3D} datacube (4825 - 5280\AA ). The curve is near-flat, and so we conclude that the SDSS and ATLAS\textsuperscript{3D} data have consistent flux calibration for this galaxy.}
	\label{sdss}
	\end{center}
\end{figure}

Next, we considered whether a single flux-calibration curve would be valid for our sample. We compared the central Mitchell spectrum for each galaxy with a SAURON spectrum obtained by summing over an equivalent region, and calculated correction curves in the same manner described above. We show the results of this process in \autoref{factors}. We find that the flux-calibration curves are broadly similar across the whole sample, and so conclude that a single calibration curve will indeed be valid for our galaxies. 

\begin{figure}
\begin{center}
\includegraphics[trim = 2cm 9cm 0cm 12cm, scale=0.8]{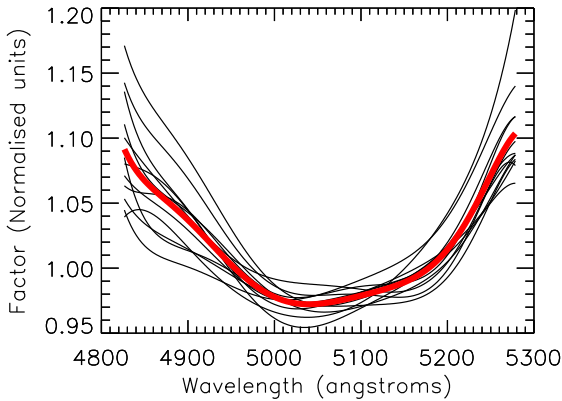}
	\caption{Mitchell-to-SAURON flux calibration curves (black lines) calculated for all galaxies in normalised units. The thick red line shows the average curve. The calculated curves are broadly similar, though a degree of scatter is evident. The root-mean square scatter from the mean is at most 4.0 per cent.}
	\label{factors}
	\end{center}
\end{figure}

We obtained an initial correction curve by comparing our central Mitchell Spectrograph spectrum for NGC 3522 with the corresponding SDSS spectrum, with the calibration curve derived in the same manner as before. We applied this curve to all Mitchell spectra, and we show the resulting Mitchell-to-SAURON curves in the left-hand window of \autoref{factorscorrected}. We find that, while the calibration is improved significantly, the ATLAS\textsuperscript{3D} and Mitchell curves still show significant offsets on the blue end. 

We therefore derive an additional correction factor by fitting a seventh-order polynomial to the \textit{average} Mitchell-SAURON calibration offset, while forcing the polynomial to approach unity at wavelengths redder than the SAURON wavelength range.  We apply this polynomial curve to the Mitchell-to-SDSS curve found previously, thus obtaining a final flux calibration curve which we apply to all spectra for our galaxies. We show the resulting Mitchell-to-SAURON calibration curves in the right-hand side of \autoref{factorscorrected}. The mean calibration curve is consistently below 1\%, with the root-mean-square scatter from the mean at most 4\% .

\begin{figure*}
\begin{center}
\includegraphics[trim = 0cm 8cm 0cm 12cm, scale=0.7]{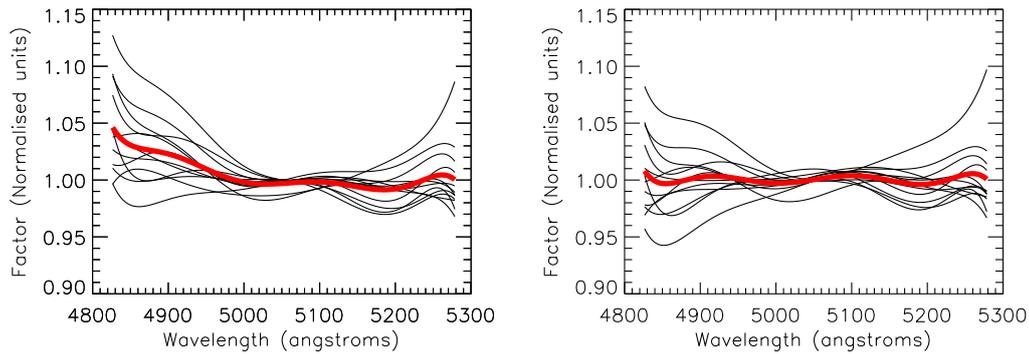}
	\caption{Mitchell-to-SAURON flux calibration curves, with all lines as in \autoref{factors}. In the left hand side, we have applied to the Mitchell data the Mitchell-to-SDSS correction curve obtained for NGC 3522. In the right hand side, we have further applied the Mitchell-to-SAURON correction curve discussed in the text. The calibration between the two instruments is significantly improved by our flux calibration procedure.}
	\label{factorscorrected}
	\end{center}
\end{figure*}

\section{Kinematics}\label{kin}

In this section, we describe the extraction of line of sight kinematics for our ETG sample. We extract stellar kinematics up to the fourth Gauss-Hermite moment, and we quantify the angular momentum of the galaxies as a function of position.  We also measure fluxes and line-of-sight kinematics of the ionised gas components, in order to clean our spectra of emission. We present our stellar kinematics in section 3.1 and we compare them with ATLAS\textsuperscript{3D} in section 3.2. We quantify the galaxies' angular momentum in section 3.3, and we present the calculation of ionised gas fluxes and kinematics in section 3.4.

\subsection{Stellar kinematics}\label{stellarkin}

We extracted stellar kinematics using the Python implementation of the publicly available penalised PiXel Fitting (pPXF) software of \citet{cappellari2004}\footnote{Available from http://purl.org/cappellari/software}, which includes the upgrade described in \citet{cappellari2017}. The pPXF routine recovers the line-of-sight velocity distribution (LOSVD) by fitting an optimised template $G_{mod}(x)$ to an observed galaxy spectrum directly in pixel space after logarithmically rebinning the spectrum in the wavelength direction. For our implementation, we added the derived sky spectra back into our galaxy spectra and then performed a second sky subtraction in pPXF, in order to improve the subtraction in the sky-dominated outskirts of our FOV \citep[e.g.][]{weijmans2009}.  The model spectra therefore take the form

\begin{equation}\label{ppxf1}
G_{mod}(x) = \sum_{k=1}^K w_k[\lagr (cx)*T_k](x) + \sum_{l=0}^L b_l\mathcal{P}_l(x) + \sum_{n=1}^N s_n S_n(x)
\end{equation} 

where $\lagr(cx)$ is the broadening function, $T_k$ a set of distinct stellar templates and $w_k$ the optimal weights of those templates, with $*$ describing convolution.  $\mathcal{P}_l(x)$ are Legendre polynomials of order $l$, with $b_l$ the corresponding weights; these are used to adjust for low-frequency differences between model and data. Likewise, $s_n$ and $S_n$ are the optimal sky weights and the sky templates themselves, respectively. For a given galaxy, the sky templates consist of all sky observations taken as part of that galaxy's observing run.

In this case, we have allowed for a tenth-degree additive Legendre polynomial correction. An alternative would be to instead use a multiplicative polynomial correction. However, such a correction is limited by the input stellar templates, and so is less free to account for low-frequency residuals when compared to an additive correction; as such, we prefer using an additive correction when deriving stellar kinematics.

The pPXF routine makes use of an input ``bias parameter", which prevents spurious solutions by biasing the recovered LOSVD towards a perfect Gaussian when $h_3$ and $h_4$ become ill-determined. It is important to select this parameter accurately: too high a value will bias the kinematics towards a Gaussian distribution even when $h_3$ and $h_4$ could be determined accurately, while too low a value risks over-fitting of noisy data. We therefore optimised the bias parameter using the simulation code included in the pPXF package, with the standard prescription that the deviation between input and output $h_3$ and $h_4$ should be less than rms/3 for values of $\sigma$ greater than three times the velocity scale. This led to an optimal bias of 0.2 for our target S/N of 20.

\begin{figure*}
\begin{center}
	\includegraphics[trim = 1cm 0cm 0cm 12cm, scale=0.85]{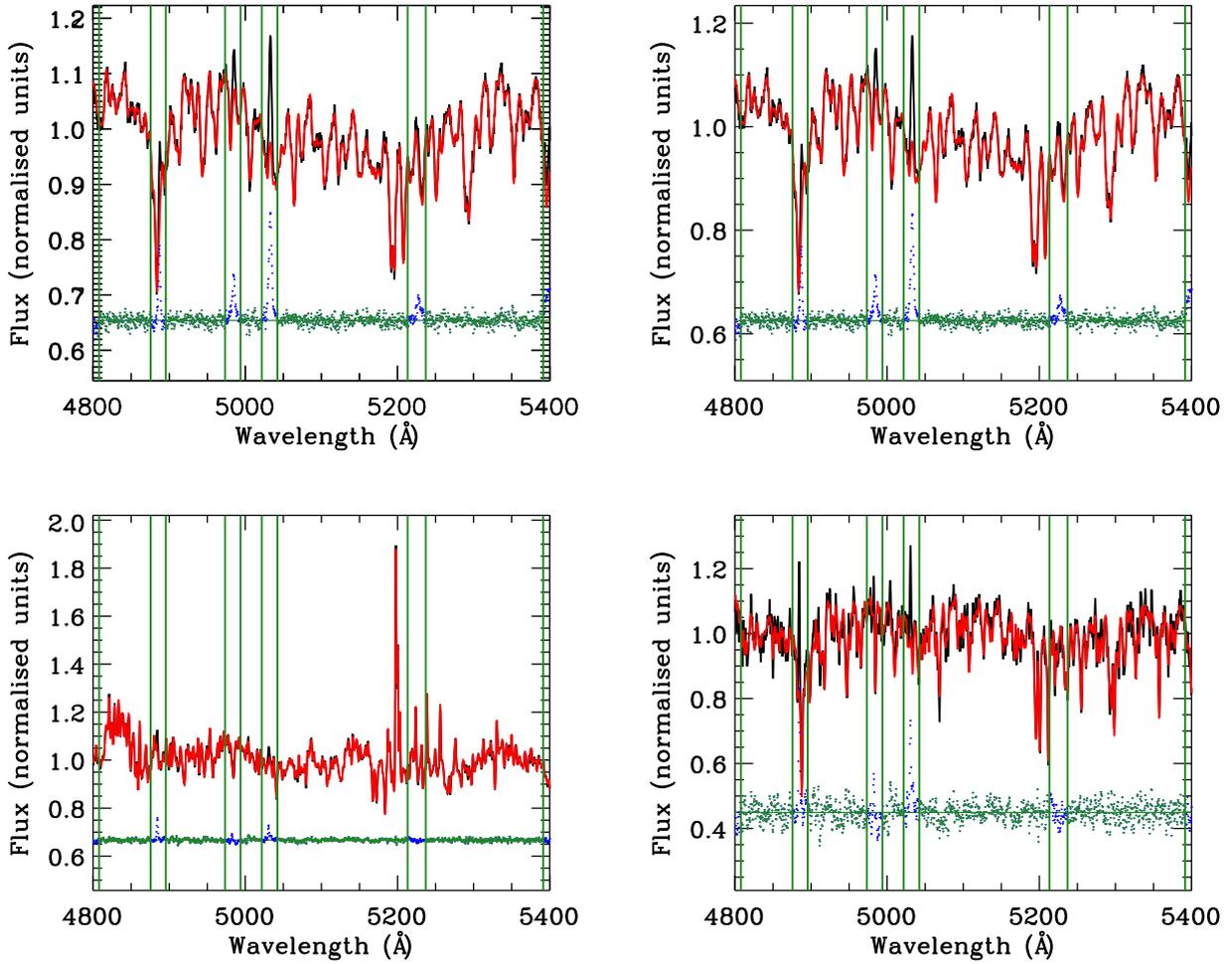}
	\caption{Example pPXF fits of high-signal (top) and sky-dominated (bottom) spectra from NGC 3626, before (left) and after (right) pPXF has been used to subtract the sky. Black lines show the observed spectra and red lines show the best-fit superposition of templates. Vertical green lines indicate regions which we excluded from the fit, green dots the fit residuals and blue dots the residuals over excluded regions. The sky provides an insignificant contribution to the top spectrum, while for the bottom spectrum pPXF is able to fit and subtract the dominant sky component.}
	\label{ppxfexample}
	\end{center}
\end{figure*}

For stellar templates, we used stellar templates from the full medium-resolution (FHWM $= 0.51$\AA)  ELODIE library \citep{prugniel2001} of observed stars. It is computationally expensive to fit the full library to each individual spectral bin; we therefore selected templates from the library by binning each galaxy into a series of elliptical annuli using the global ellipticities and position angles derived in \citet{kraj2011} from SDSS and INT imaging data; we performed pPXF fits on these annuli using the full ELODIE library, and we selected any star which was given a non-zero weight. We then ran pPXF over all galaxy spectra with the emission line regions (H$\beta$, [\OIII ] and [\NI ]) masked out, while iteratively detecting and masking bad pixels; we also masked the red and blue edges of all spectra when fitting, to avoid any potential problems relating to flat-fielding. We show some example pPXF fits in \autoref{ppxfexample}, and we present the resulting line-of-sight kinematics in Figures 7-10.

\begin{figure*}
\begin{center}
	\includegraphics[trim = 1cm 0cm 0cm 2cm, scale=0.85]{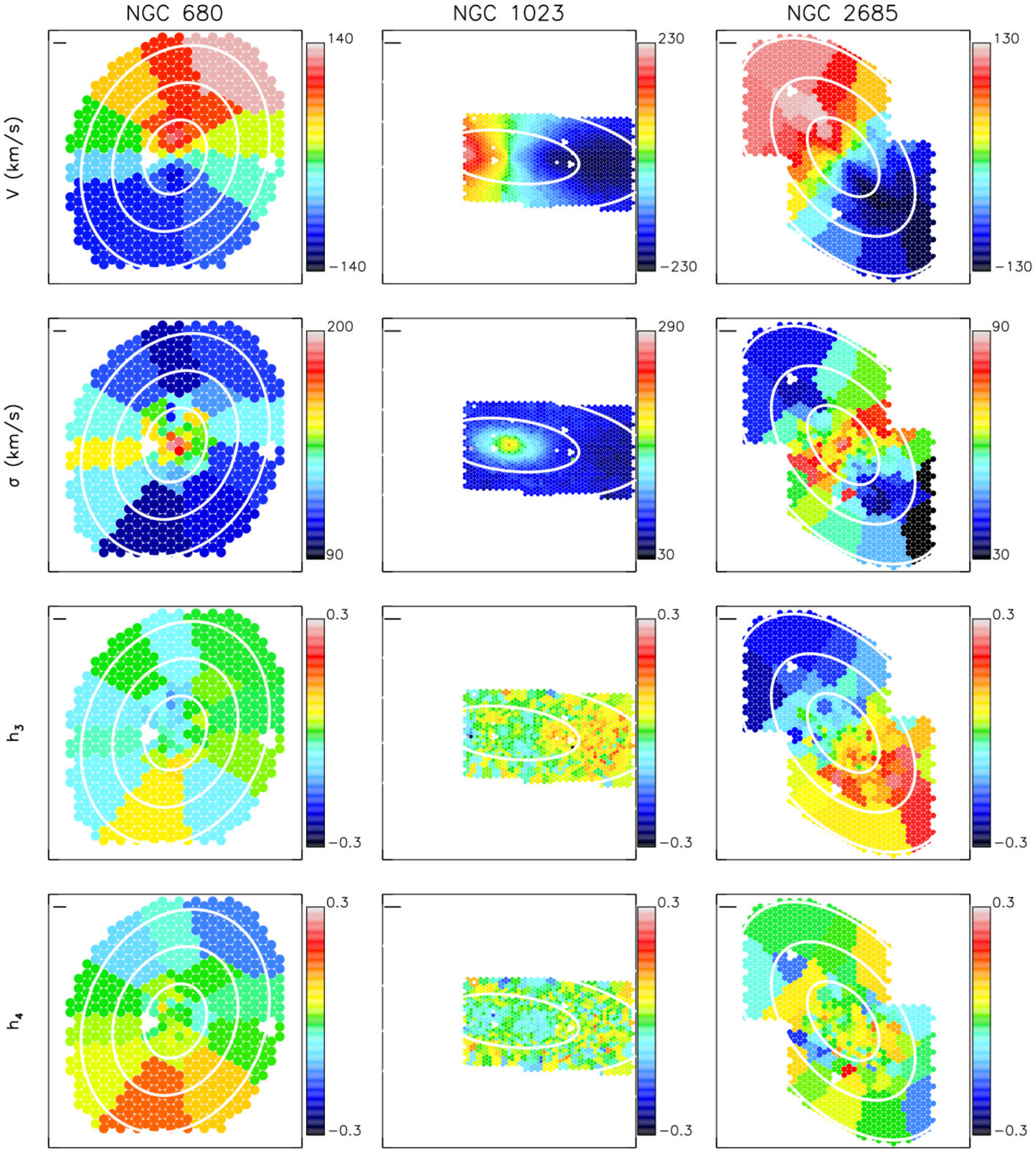}
	\caption{Kinematics maps for NGC 680, NGC 1023 and NGC 2685. Rows from top to bottom: velocity (km/s), velocity dispersion (km/s), $h_3$, $h_4$. The solid black lines mark a length of 1 kpc. The white contours are spaced in units of $R_e$. Fibre positions from the missing NGC 2685 dither on the top left of our FOV have been re-added for presentational purposes, with kinematics assigned to each from the nearest Voronoi bin.}
	\label{vels}
	\end{center}
\end{figure*}

\begin{figure*}
\begin{center}
	\includegraphics[trim = 1cm 0cm 0cm 2cm, scale=0.85]{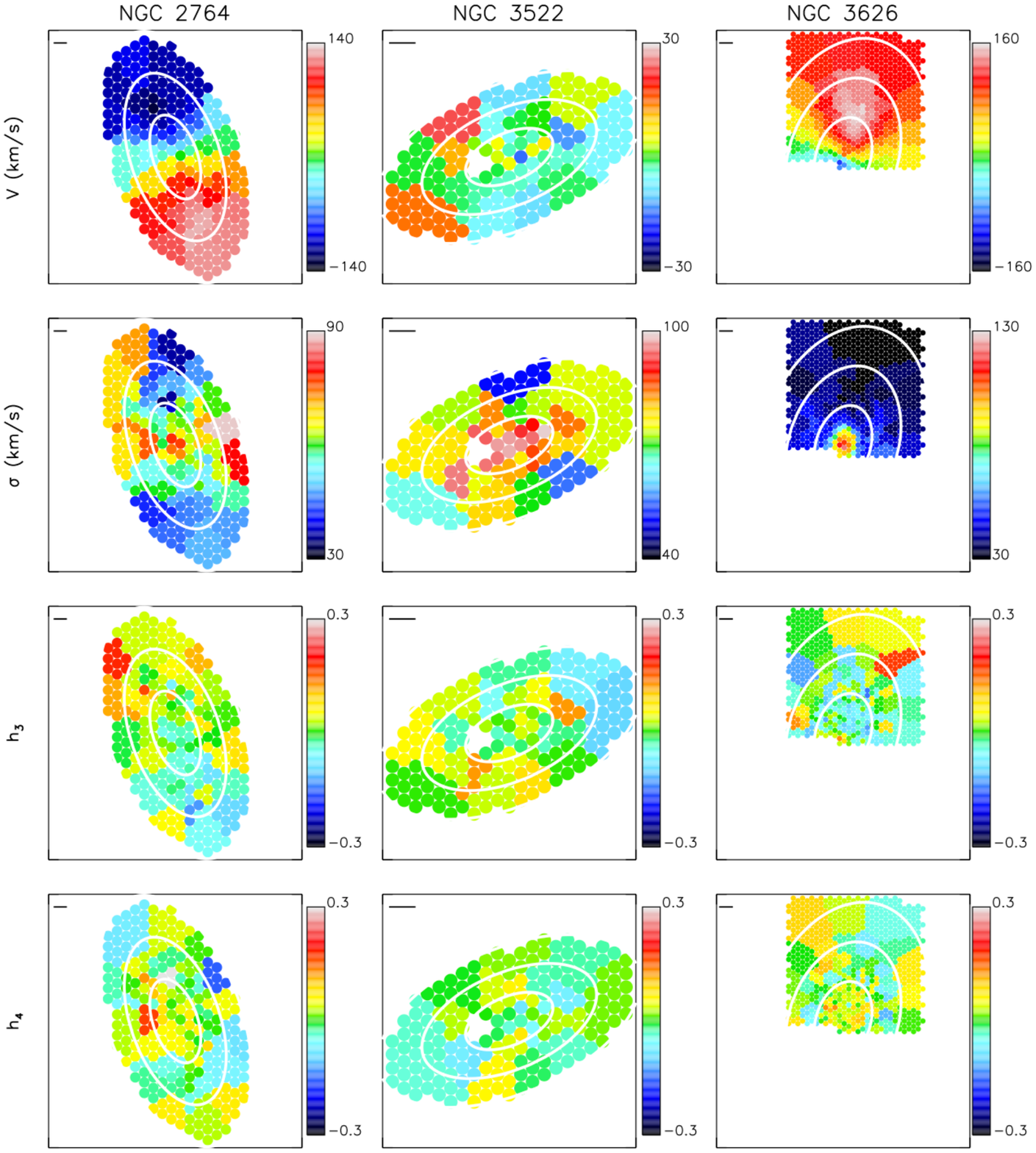}
	\caption{As in \autoref{vels}, but for galaxies NGC 2764, NGC 3522 and NGC 3626.}
	\label{dispersions}
	\end{center}
\end{figure*}

\begin{figure*}
\begin{center}
	\includegraphics[trim = 1cm 0cm 0cm 2cm, scale=0.85]{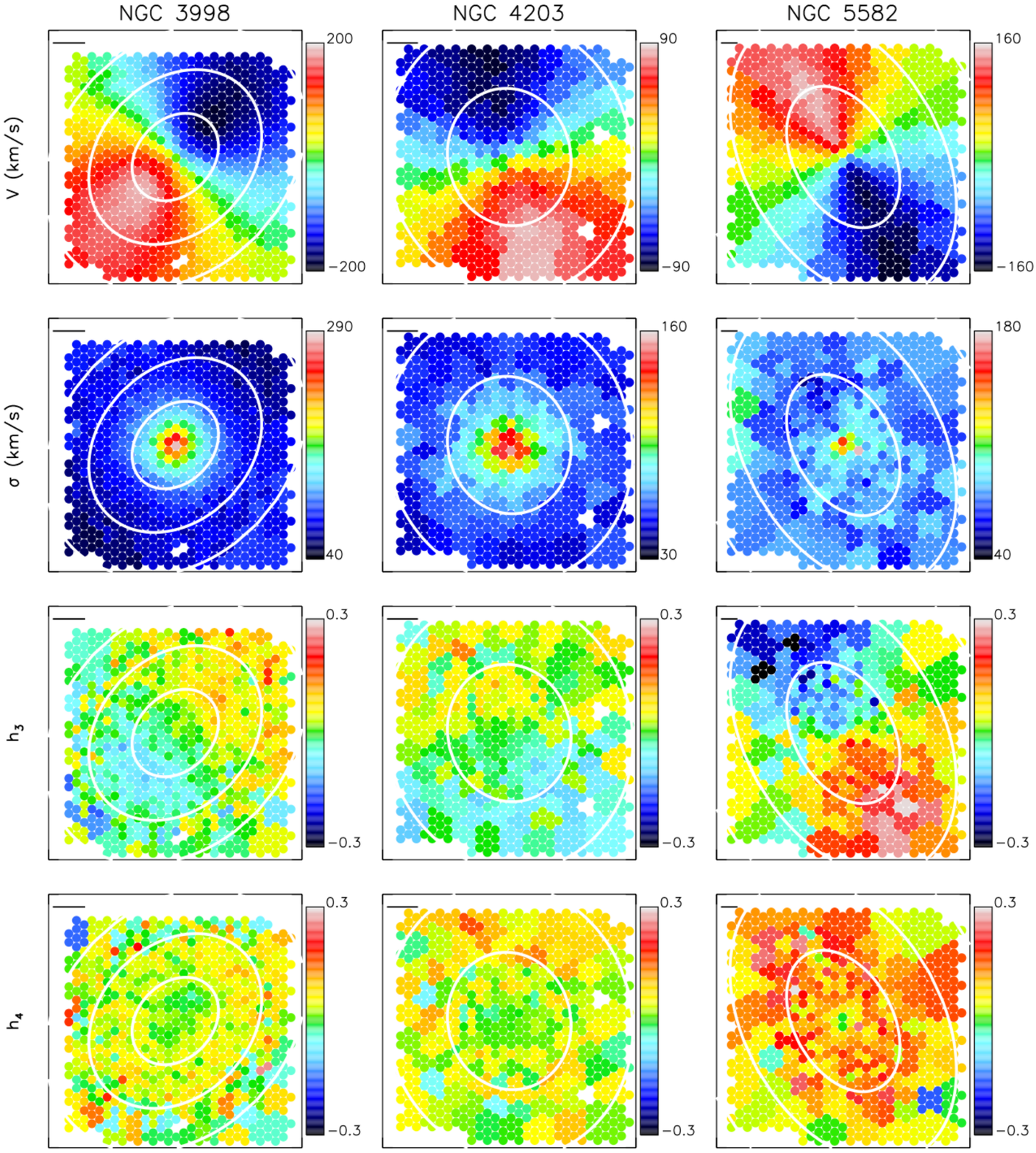}
	\caption{As in \autoref{vels}, but for galaxies NGC 3998, NGC 4203 and NGC 5582. Fibre positions from the missing NGC 4203 dither have been re-added for presentational purposes, with kinematics assigned to each from the nearest Voronoi bin.}
	\label{h3s}
	\end{center}
\end{figure*}

\begin{figure*}
\begin{center}
	\includegraphics[trim = 1cm 0cm 0cm 2cm, scale=0.85]{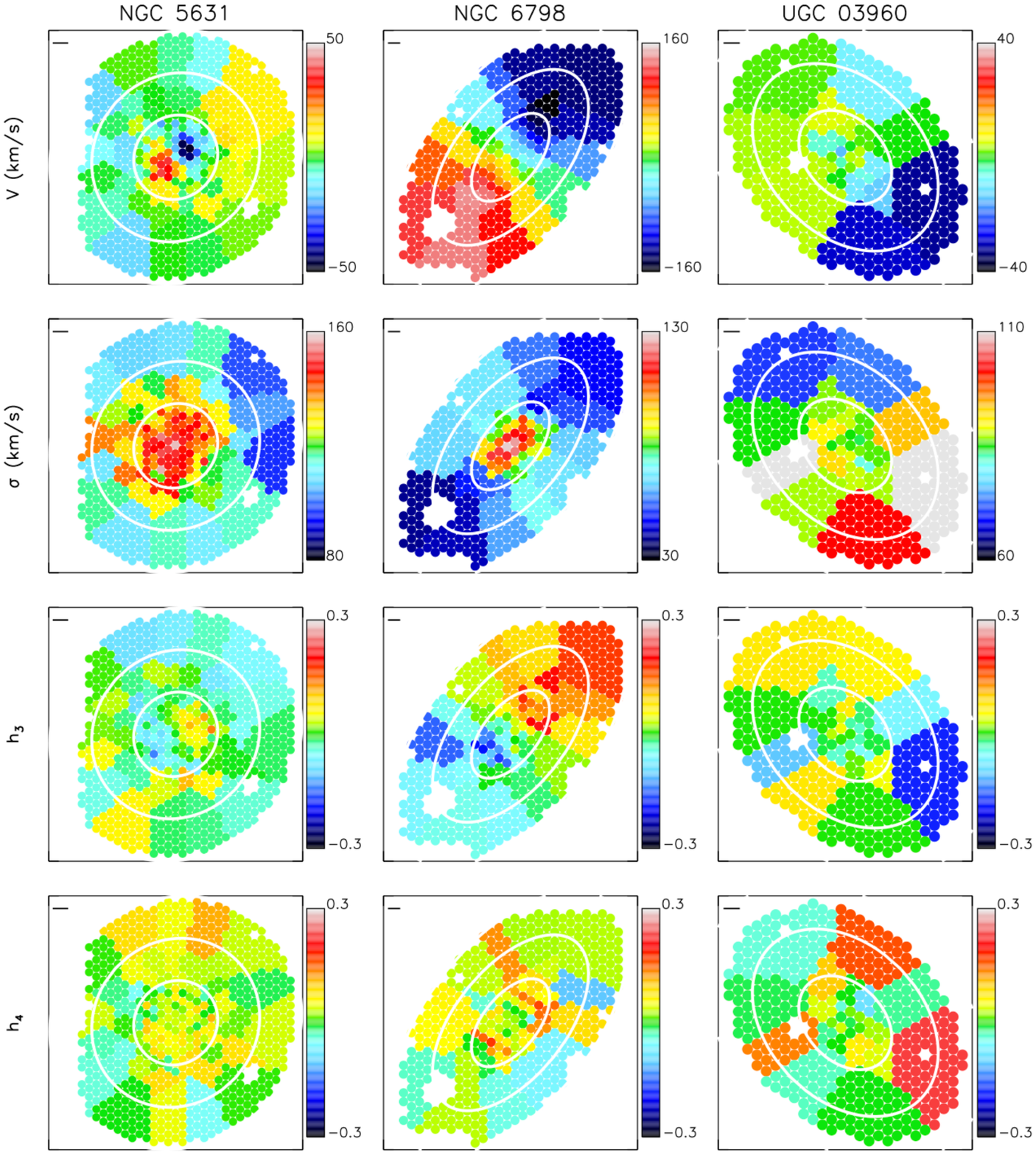}
	\caption{As in \autoref{vels}, but for galaxies NGC 5631, NGC 6798 and UGC 03960.}
	\label{h4s}
	\end{center}
\end{figure*}

We determined random measurement errors by adding Gaussian noise to the spectra and rerunning the fits with zero bias for 100 iterations each. However, these errors alone do not consider systematic effects - such as template mismatch and imperfect sky subtraction - and so will underestimate the true level of uncertainty \citep[e.g.][]{arnold2014}. We estimate the level of systematic error in a similar manner to \citet{boardman2016}. We derive pPXF kinematics using the MILES library of observed stars \citet{sanchezblazquez2006} in the same manner as above, after broadening our spectra to match the MILES library resolution of 2.51 \AA\ \citep{falconbarroso2011}. We compute the residuals between the derived kinematics, and we then calculate the $1\sigma$ dispersion between the residuals with respect to zero. We derive systematic error values of 3.3 km/s, 4.3 km/s, 0.03 and 0.04 for the velocity, dispersion, $h_3$ and $h_4$ respectively, which we add in quadrature to the original errors; this results in median overall error values of 4.1 km/s, 5.7 km/s, 0.04 and 0.05 respectively.

We compare the kinematics from ELODIE and MILES templates in \autoref{libcompare}. Our findings are essentially identical to what was previously reported for NGC 3998 in \citet{boardman2016}: we find 1-1 agreement in the velocity, $h_3$ and $h_4$ values - albeit with non-negligible levels of scatter - but we find that the MILES results tend towards higher dispersion values when the ELODIE velocity dispersion is low. We view the ELODIE results as being more reliable in such cases due to the MILES library's higher intrinsic dispersion of $\sim 60$ km/s, which is above what we find with ELODIE in most of the galaxies' outskirts. For ELODIE dispersions below 60km/s, the derived MILES dispersion is $5.5$km/s higher on average.

\begin{figure}
\begin{center}
	\includegraphics[trim = 2cm 4cm 0cm 7cm, scale=0.45]{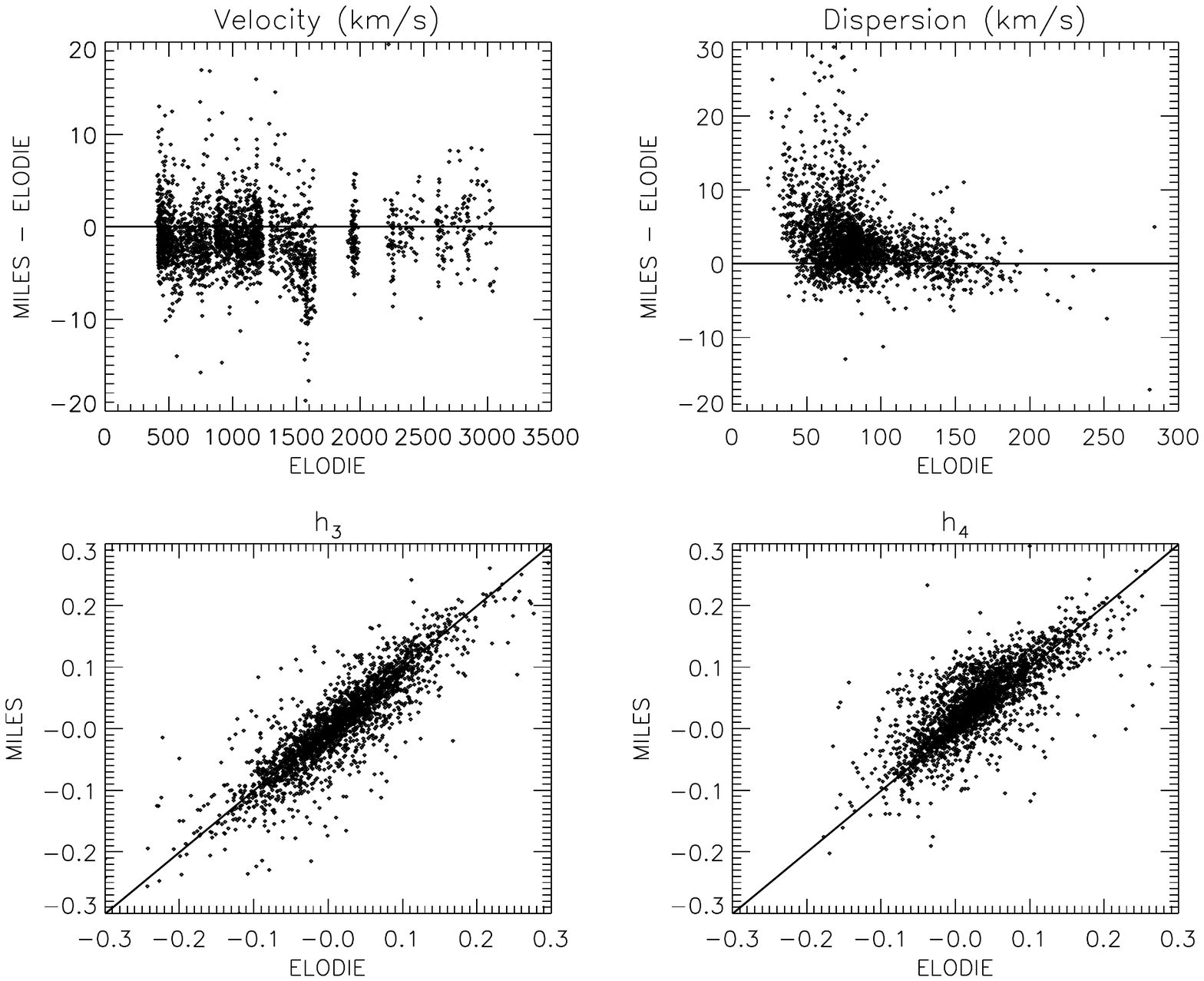}
	\caption{Comparison of Mitchell kinematics for our twelve galaxies derived using ELODIE and MILES libraries in pPXF. We show absolute values of $h_3$ and $h_4$ from both ELODIE and MILES libraries, while we show the relative differences between ELODIE and MILES velocities and dispersions in order to emphasise differences. We find good overall agreement in the velocity, $h_3$ and $h_4$ values, though with non-negligible scatter. We also find that the MILES dispersion values approach larger values when the ELODIE value is low, which we argue to be due to the MILES library's higher intrinsic dispersion.}
	\label{libcompare}
	\end{center}
\end{figure}

We assessed the level of systematic error due to imperfect sky subtraction, by comparing the kinematics described above to two sets of kinematics obtained after over-subtracting and under-subtracting the sky by 10\% respectively. For each of these new kinematic datasets, we compare to the original kinematic dataset and then calculated systematic errors in the same manner described previously. This led to maximum systematic error values of 1.7km/s, 1.7km/s, 0.02 and 0.02 for velocity, dispersion, $h_3$ and $h_4$ respectively; these values are small compared to the sources of error already considered, so we do not factor these into our error calculation. 

We also derived pPXF kinematics for the first two moments $(V,\sigma)$ only, for use in cases where the higher-order moments are not required; this is to prevent any dependence on $h_3$, $h_4$ or the pPXF penalty parameter in cases where only the first two moments are needed. We obtained errors in the same way as before, deriving systematic error terms of 5.9 km/s and 8.2 km/s for the velocity and velocity dispersion in turn; this produces median overall error values of 6.5 km/s and 9.1 km/s.

\subsection{Comparison with ATLAS\textsuperscript{3D}}\label{atlas3d}%

As all of our galaxies form part of the ATLAS\textsuperscript{3D} survey, it is useful to compare our kinematics with those derived previously with the SAURON instrument. We first perform this comparison by finding all Mitchell spectra for which an ATLAS\textsuperscript{3D} bin lies within the radius of a Mitchell fibre (2.08$''$); we then compare the kinematics of these bins with those of the closest ATLAS\textsuperscript{3D} bin. We compare the velocities and dispersions calculated from 2-moment pPXF fits, as well as the kinematics calculated from 4-moment fits. We found excellent agreement in the line-of-sight velocities calculated between the two datasets; however, we also found the ATLAS\textsuperscript{3D} velocity dispersions to be systematically higher on average when fitting for $h_3$ and $h_4$, while the $h_3$ and $h_4$ comparisons show a high degree of scatter.

There are several potential confounding factors in this comparison, however. The ATLAS\textsuperscript{3D} spectra for our galaxies have an instrumental dispersion of $\sigma_{\rm instr}\approx98$ km/s, and we find many galaxies to have dispersions below this even within the area covered by ATLAS\textsuperscript{3D}. In addition, the ATLAS\textsuperscript{3D} datacubes are often significantly binned towards the edge of their FOV, which could serve to artificially enhance the measured dispersion \citep[e.g.][]{arnold2014}. Lastly, the Mitchell fibres are significantly larger than the 0.8$''$ $\times$ 0.8$''$ spaxels employed by ATLAS\textsuperscript{3D}.

\begin{figure}
\begin{center}
	\includegraphics[trim = 1cm 1cm 1cm 15cm, scale = 0.9]{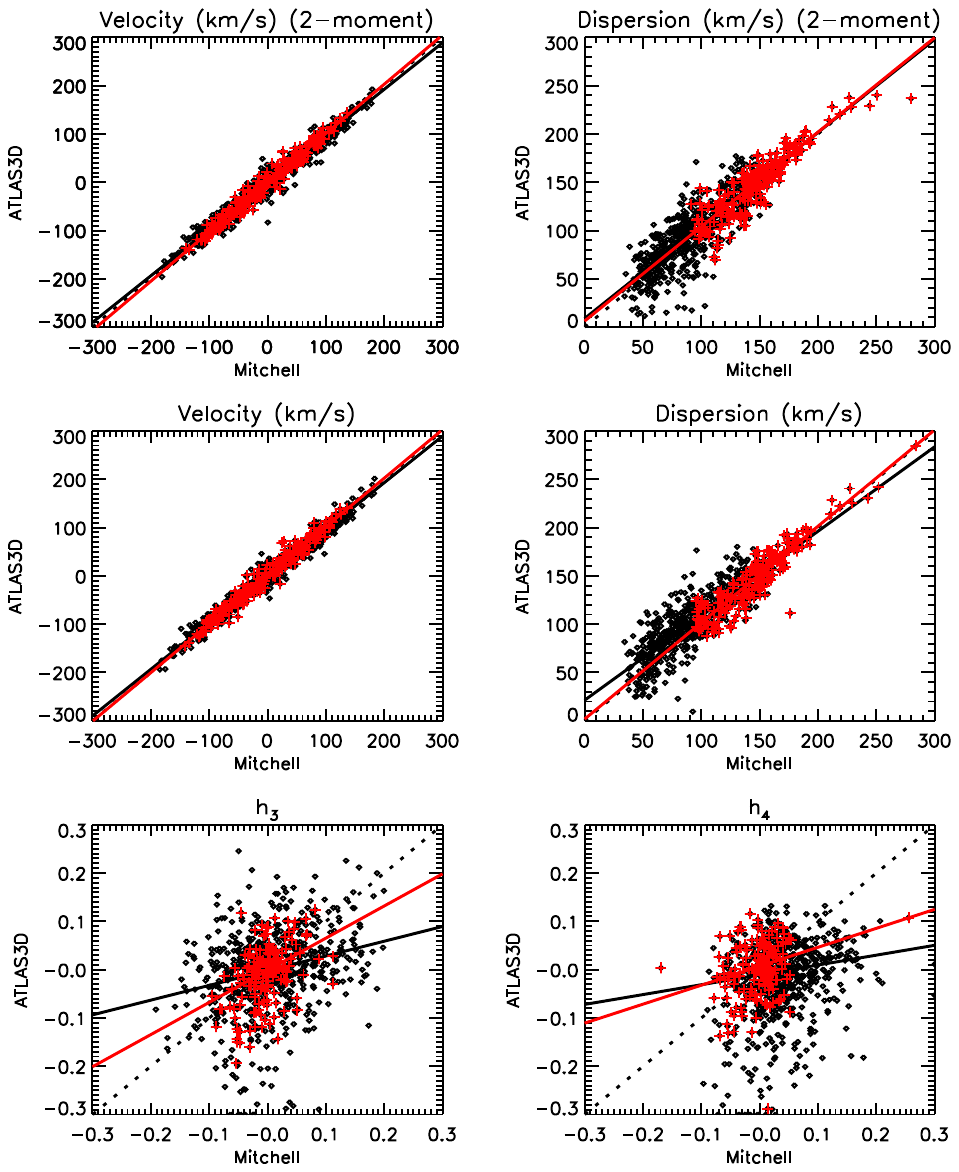}
	\caption{Comparison of the Mitchell kinematics of our twelve galaxies with the kinematics reported in ATLAS\textsuperscript{3D}. The dotted lines show the one-to-one relation, while the solid lines are obtained from a robust least-absolute-deviation fit. Unbinned high-dispersion datapoints are highlighted with red crosses; the red solid lines show robust least-absolute-deviation fits to these points. We find that significantly improved consistency with ATLAS\textsuperscript{3D} once low-dispersion and/or binned datapoints are excluded. }
	\label{atlas2}
	\end{center}
\end{figure}

Motivated by the above, we performed a second comparison in the following way. We first selected datapoints to compare in the same manner as before. We then excluded any datapoints for which the Mitchell Spectrograph velocity dispersion was below 98 km/s, and we further excluded datapoints for $h_3$ and $h_4$ for which the Mitchell dispersion was below 120 km/s; this is to ensure that the dispersion could be accurately measured in ATLAS\textsuperscript{3D} without strong penalisation of $h_3$ or $h_4$. We further limited ourselves to data points in which neither the Mitchell nor the ATLAS\textsuperscript{3D} spectra had been binned, in order to ensure that the velocity dispersion in one or both datasets was not being enhanced by binning. We show the results of this in \autoref{atlas2}, along with the results of the first comparison discussed previously. We now find excellent agreement between the velocity and dispersion of the two datasets, though the $h_3$ and $h_4$ comparisons continue to show a large degree of scatter.

\begin{figure}
\begin{center}
	\includegraphics[trim = 2cm 1cm 1cm 15cm, scale=0.9]{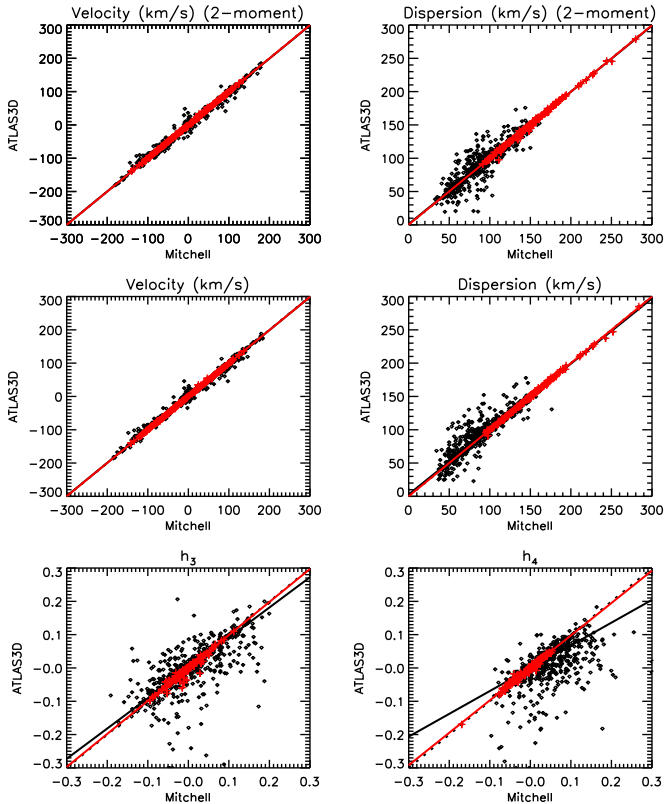}
 \caption{As in \autoref{atlas2}, except that the Mitchell datapoints have been matched to the SAURON datapoint within 2.08 arcseconds that is \textit{closest in value}. All lines and symbols are as before. We find one-to-one agreement once low-dispersion and binned datapoints have been removed, with very little scatter.} 
	\label{atlas3}
	\end{center}
\end{figure}

We present one final comparison in \autoref{atlas3}. Here, we take each Mitchell kinematic datapoint in turn and find all SAURON datapoints within the Mitchell fibre radius, and then compare each Mitchell datapoint to the selected SAURON datapoint that is \textit{closest in value} to the Mitchell datapoint being considered. Unbinned high-dispersion datapoints were selected in the same manner as described previously. We find tight one-to-one relations with almost no scatter in this case, once datapoint affected by binning or low dispersions have been excluded. We therefore find our data to be fully consistent with that from the ATLAS\textsuperscript{3D} survey.

\subsection{Stellar angular momentum}

We tested our stellar kinematics for transitions beyond 1$R_e$, by considering the galaxy angular momentum as a function of radius. We explore the angular momentum of the galaxies by using the $\lambda_R$ parameter \citep{emsellem2007} as proxy, which in the case of two-dimensional spectroscopy takes the form 

\begin{equation}\label{lit1}
\lambda_R = \frac{\langle \sum_{i=1}^{N_P} F_i R_i |V_i|\rangle}{\langle R \sum_{i=1}^{N_P} F_i R_i \sqrt{V_i^2 + \sigma_i^2}\rangle}
\end{equation}

\noindent

where $R$ signifies the mean radius of an ellipse, $F$ represents the flux, $V$ is the line-of-sight velocity and $\sigma$ the velocity dispersion. We calculate $\lambda_R$, by summing over all fibres for which the centre falls within a given ellipse, after applying to each fibre the kinematics of its corresponding Voronoi bin. We do not attempt to calculate the profiles beyond $R = R_{\rm max}$, to ensure that the ellipse is always well-filled with spectra. We used velocities and dispersions from the pPXF fits to $V$ and $\sigma$ only. We symmetrized the kinematics and fibre positions when calculating $\lambda_R$ for NCG 1023 and NGC 3626, replicating the kinematics values on the missing sides, due to these galaxies not being centred on our FOV. We used ellipses of constant ellipticity, based on the global ellipticities and position angles for these galaxies reported in \citet{kraj2011}.

\begin{figure}
\begin{center}
	\includegraphics[trim = 2cm 13cm 4cm 4cm, scale=0.7]{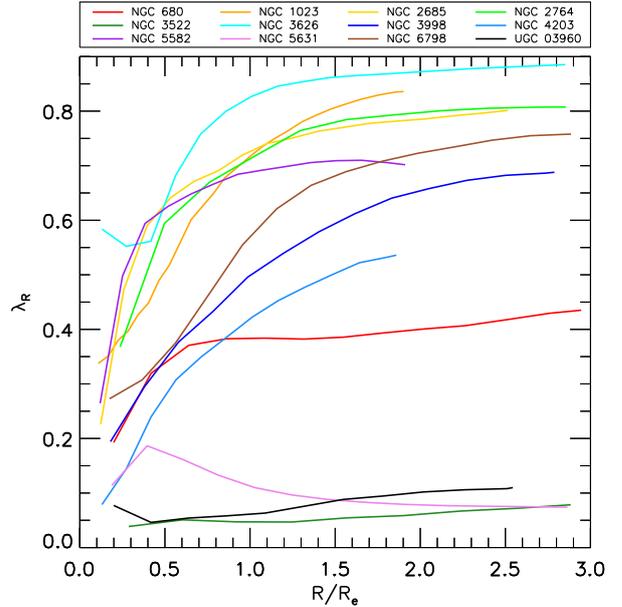}
	\caption{$\lambda_R$ profiles for our twelve ETGs. We find no abrupt drops beyond the central half-light radius, with all but one galaxy (NGC 5631) showing rising profiles overall.}
	\label{lamr}
	\end{center}
\end{figure}

We present $\lambda_R$ profiles for our sample in \autoref{lamr}.  We find no clear transitions in the galaxies' angular momentum beyond the central effective radius: beyond $1R_e$, galaxies in our sample show the expected slightly rising  or rather constant $\lambda_R$ profiles. Hence, galaxies that are FRs (SRs) at $1R_e$, following the \citet{emsellem2011} definitions, keep their high (low) $\lambda_R$ profiles over the full FOV relative to their respective ellipticities. All but one of our galaxies show slightly rising $\lambda_R$ profiles overall; the exception here is NGC 5631, which is a known kinematically-decoupled core (KDC), with isophotes becoming nearly round at large radii and with an associated expected decline in its $\lambda_R$ profile.

\subsection{Ionised gas kinematics}\label{gas}

We extracted ionised gas fluxes and kinematics using the GANDALF code of \citep{sarzi2006}, with the aim of cleaning our spectra of emission.

Previously, we ran pPXF allowing for an \textit{additive} Legendre polynomial correction. Such a correction can produce unrealistic features in model spectra over masked wavelength regions, due to the correction's lack of dependence on the stellar templates; this makes an additive correction ill-suited for extracting ionised gas emission lines. We therefore reran pPXF allowing for a 10th degree \textit{multiplicative} Legendre polynomial instead, which accounts for residual continuum variation while avoiding degeneracies with individual line strengths. We again used pPXF to fit and subtract the sky, therefore obtaining model galaxy spectra of the form 

\begin{equation}\label{ppxf2}
G_{mod}(x) = \sum_{k=1}^K w_k[\lagr (cx)*T_k](x) \times \sum_{l=1}^L b_l\mathcal{P}_l(x)  + \sum_{n=1}^N s_n S_n(x)
\end{equation}  

\noindent in which all symbols are as before. We derive systematic error terms by recalculating the kinematics with MILES stars as before, finding values of 3.1 km/s, 4.7 km/s,0.03 and 0.04 for each kinematic moment in turn.

The GANDALF code uses the pPXF-derived stellar kinematics as inputs in order to derive a new optimal stellar template along with an accompanying optimal emission template.  Flux values are calculated for each emission line, along with the first two kinematic moments $(V,\sigma)$. 

We implemented GANDALF as follows.  We again allowed for a tenth-order multiplicative Legendre polynomial correction, to account for continuum variation. We searched for three ionised gas features in each of our spectra: H$\beta$ line, the [\OIII] doublet and the [\NI] doublet. We performed an initial GANDALF fit for each binned spectrum in which the H$\beta$ and [NI] emission regions were masked, to derive kinematics for the [\OIII] emission; afterwards, we fit for all expected emission features with the gas kinematics fixed to the [\OIII] value. We calculated for each spectrum the amplitude-over-noise (A/N) of all detected features in the initial fit. We then extracted all [\OIII] features with $A/N > 4$ and we extracted all H$\beta$ and [\NI] features with $A/N > 3$, following the reasoning in \citet{sarzi2006}. We present an example GANDALF fit in \autoref{gandalfexample}.  We applied the GANDALF-derived continuum correction to each of our spectra before proceeding, in order to account for any residual continuum contamination.

\begin{figure}
\begin{center}
	\includegraphics[trim = 1cm 0cm 0cm 12cm, scale=0.65]{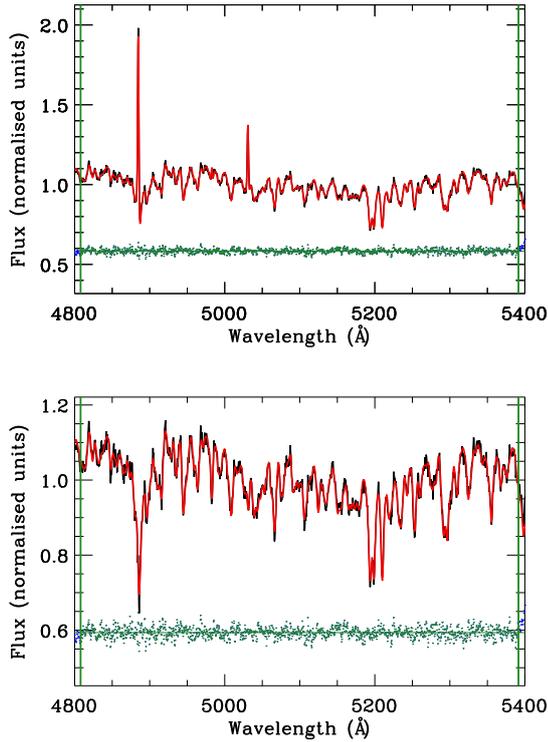}
	\caption{Example GANDALF fit to a sky-subtracted galaxy spectrum from NGC 3626, before (top) and after (bottom) subtracting the GANDALF-derived emission component. All lines are as in \autoref{ppxfexample}}
	\label{gandalfexample}
	\end{center}
\end{figure}

Although the above procedure produced good fits for the vast majority of spectra in our sample, we noted significant fit residuals around the H$\beta$ region over much of the FOV for NGC 2764. Given that H$\beta$ is the only age-sensitive stellar absorption feature within the wavelength range of our data, we exclude NGC 2764 from the stellar population analysis described in \autoref{popsppxf}.

We present our extracted [\OIII] fluxes and velocities in \autoref{gasflux} and \autoref{gasvel}. We detect ionised gas well beyond the central effective radius for many of our galaxies. We observe several cases in which the ionised gas is counter-rotating or else misaligned with respect to the stars; we demonstrate this in \autoref{patable}, in which we calculate the misalignment angles using the method described in Appendix C of \citet{kraj2006} \footnote{Available from http://purl.org/cappellari/software}. This latter point is not a new result for these systems, being already apparent from the SAURON gas maps published by ATLAS\textsuperscript{3D} \citep{davis2011}, but is relevant when considering these galaxies' evolutionary pasts. 

\begin{table}
\begin{center}
\begin{tabular}{|c|c|c|c|}
\hline 
Galaxy & $\theta_{*}$  & $\theta_{ion}$ & $|\theta_{*} - \theta_{ion}|$ ($^{\circ}$) \\ 
\hline 
NGC 680 & $359.5 \pm 8.5$ & $22.5 \pm 10.3$ & 23.0 \\ 
\hline 
NGC 1023 & $86.5 \pm 1.5$ & N/A & N/A \\ 
\hline 
NGC 2685 & $38.0 \pm 3.8$ & $88.5 \pm 5.5$  & 50.5 \\ 
\hline 
NGC 2764 & $196.5 \pm 7.8$ & $185.5 \pm 13.0$ & 9.0 \\ 
\hline 
NGC 3522 & $116.5 \pm 89.8$ & $181.0 \pm 24.5$ & 64.5 \\ 
\hline 
NGC 3626 & $343.0 \pm 4.0$ & $168.0 \pm 4.8$ & 175.0 \\ 
\hline 
NGC 3998 & $136.5 \pm 2.0$ & $87.5 \pm 2.5$  & 49.0 \\ 
\hline 
NGC 4203 & $194.5 \pm 6.8$ & $198.0 \pm 5.8$  & 3.5 \\ 
\hline 
NGC 5582 & $29.0 \pm 3.3$ & $30.5 \pm 3.5$ & 1.5 \\ 
\hline 
NGC 5631 & $132.5 \pm 31.5$ & $319.5 \pm 6.5$ & 173.0 \\ 
\hline 
NGC 6798 & $139.0 \pm 7.8$ & $310.5 \pm 7.0$ & 171.5 \\ 
\hline 
UGC 03960 & $33.5 \pm 89.8$ & $97.0 \pm 39.8$ & 57.2  \\ 
\hline
\end{tabular} 
\end{center}
\caption{Stellar and ionised gas kinematic position angles and uncertainties of our ETG sample, measured anticlockwise from north for the receding part of the velocity map using the method of \citet{kraj2006}, along with the misalignment angles between the two. We do not report ionised gas position angle for NGC 1023 due to this galaxy's higher irregular gas kinematics.}
\label{patable}
\end{table}

\begin{figure*}
\begin{center}
	\includegraphics[trim = 1cm 3cm 0cm 2cm, scale=0.75]{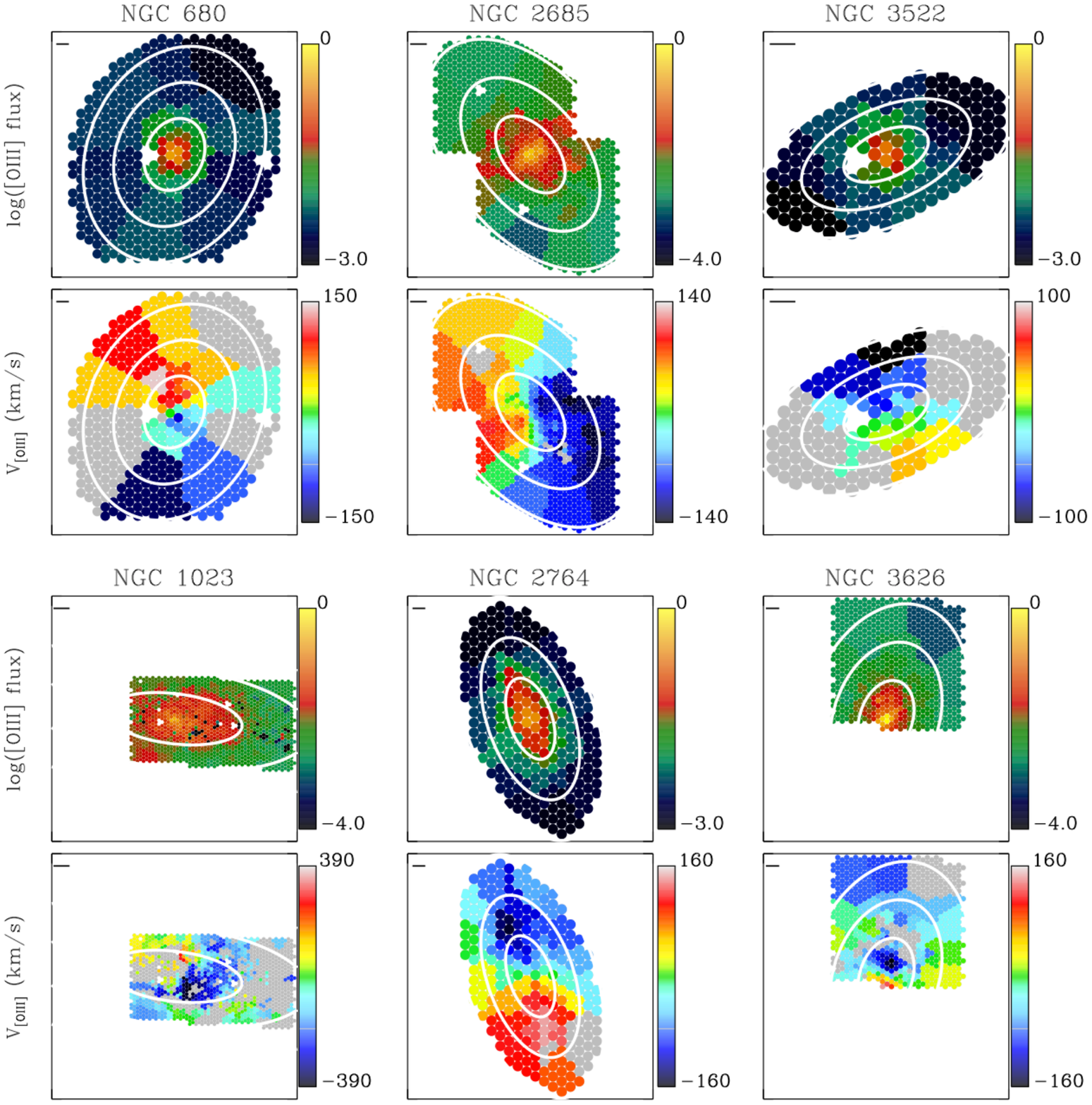}
	\caption{Maps of log(\OIII\ flux) (top, arbitrary units) and \OIII\ velocity (bottom, km/s) for the first six galaxies in the ETG sample. Flux is in arbitrary units, and has been divided through by the number of fibres comprising each spectral bin. Grey bins in the velocity maps indicate regions below our amplitude-over-noise threshold, as discussed in the text. The white contours are spaced in units of $R_e$. The solid black lines in the top left corner mark a length of 1kpc. Fibre positions from the missing NGC 2685 dither on the top left of our FOV have been re-added for presentational purposes, with kinematics and fluxes assigned to each from the nearest Voronoi bin.}
	\label{gasflux}
	\end{center}
\end{figure*}

\begin{figure*}
\begin{center}
	\includegraphics[trim = 1cm 3cm 0cm 2cm, scale=0.75]{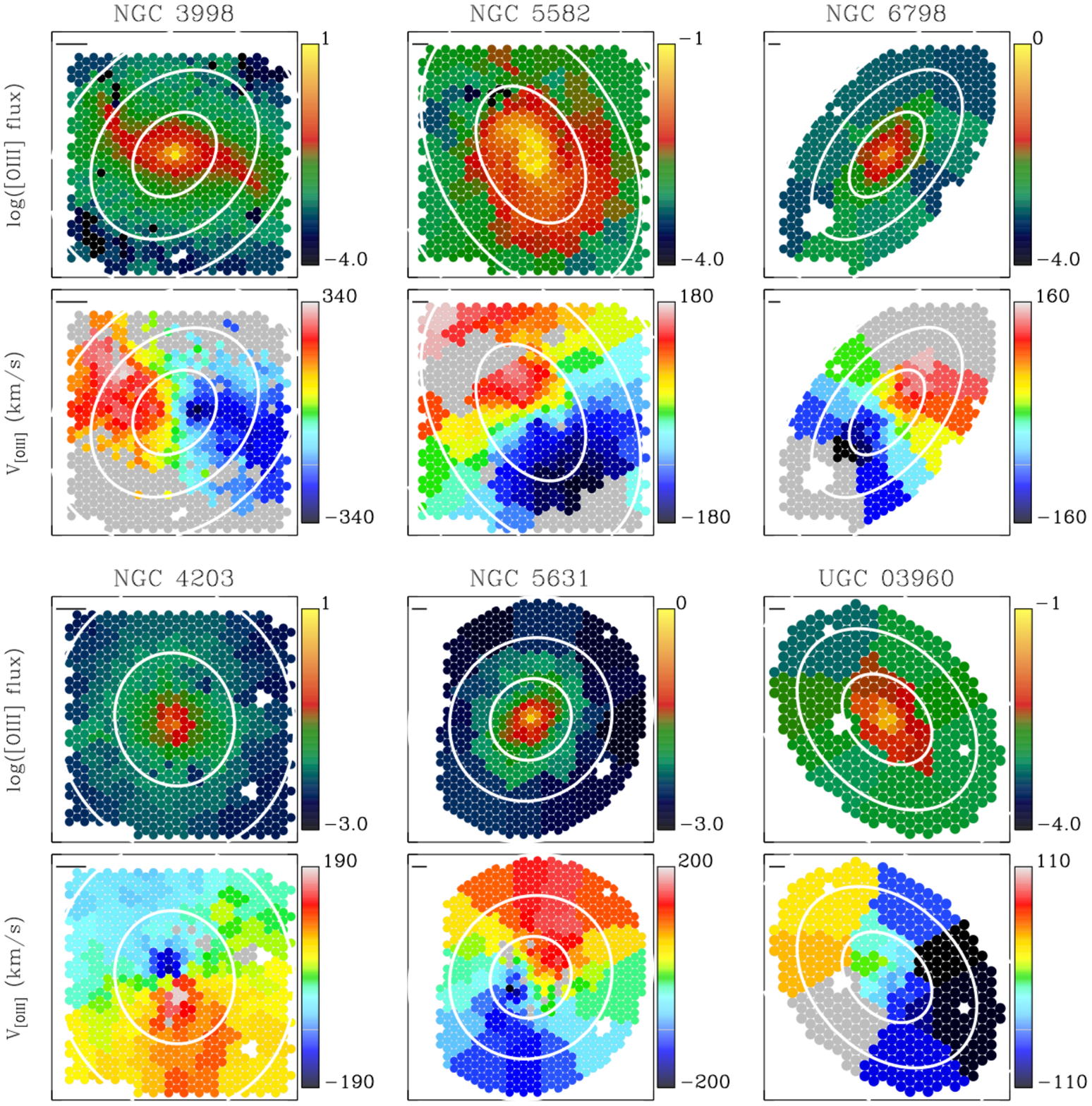}
	\caption{As in \autoref{gasflux}, for the remaining ETGs in the sample. Fibre positions from the missing NGC 4203 dither have been re-added for presentational purposes, with kinematics and fluxes assigned to each from the nearest Voronoi bin.}
	\label{gasvel}
	\end{center}
\end{figure*}

\section{Line strengths}\label{linestrength}

In this section we describe the extraction of absorption line strengths for our galaxy sample, which we will then use to constrain the stellar ages, metallicities and abundance ratios of these galaxies. Each index consists of a central bandpass, where the absorption feature is located, along with a pair of blue and red pseudo-continua. An index is measured by determining the mean of each pseudo-continuum and then drawing a straight line between the two midpoints; the index is then given by the difference in flux between the central bandpass and the line \citep{worthey1994}.

We begin by broadening all of our spectra to the 14\AA\ line-index system (LIS) proposed in \citet{vazdekis2010}, in which spectra are broadened to a constant FWHM of 14\AA . We take both the instrumental dispersion and stellar velocity dispersion into account when broadening, resulting in spectra with a FWHM of 14\AA\ independent of wavelength. We then calculate line indices for each of our spectra, using the pPXF-derived stellar velocities to determine the required amount of redshifting. We calculate the H$\beta$, Fe5015, Mg\,{\it b} and Fe5270 indices for all galaxies. We also calculate the Fe5335 index where possible; we are not able to calculate this index for all objects, as its red continuum region is redshifted beyond the available wavelength range in certain cases. 

We calculate errors by performing 500 monte-carlo re-simulations, with Gaussian noise added to the spectra and to the input pPXF line-of-sight velocities. We set lower limits on the errors by calculating the differences between the indices from a given spectrum and those from the associated optimal template, likewise broadened to 14\AA\ FWHM; in practice, this means that the errors in the inner parts of each galaxy are set by data-template differences, whereas further out we find random errors to dominate. We present two-dimensional maps of the line-indices for our sample in four supplementary figures available online. 

For visualisation purposes, we construct smoothed line index profiles as follows. We place over each galaxy a central elliptical aperture and three elliptical annuli; the central apertures have a major axis radius of 0.5$R_e$, while the annuli have major-axis boundaries at 0.5$R_e$, 1$R_e$, 1.5$R_e$ and 2.5$R_e$. We discounted the outer annulus for galaxies with $R_{\rm max} < 2.5$, to ensure good coverage for all annuli. We obtain the associated line index value across all of the resulting regions by taking the mean of all Voronoi bins whose light-weighted centres fall within each annulus. 

We present the resulting H$\beta$ and Mg\,{\it b} profiles in \autoref{lsprofiles}, in which radii are given as flux-weighted average radii for a given aperture or annulus. We find a wide range of H$\beta$ behaviours: the majority of our galaxies show flat H$\beta$ profiles, but we also find H$\beta$ to rise or fall in individual systems. We observe steep H$\beta$ rises for the galaxies NGC 2685 and NGC 6798, while we see significant falls for the galaxies NGC 2764 and NGC 3626. We find Mg\,{\it b} to fall with radius for most of our galaxies; the two exceptions here are NGC 2764 and NGC 3626, for which the Mg\,{\it b} gradient is near-flat. 

\begin{figure}
\begin{center}
	\includegraphics[trim = 2cm 5cm 0cm 12cm, scale=0.45]{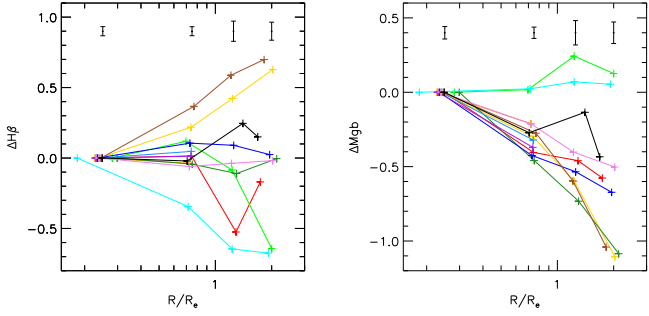}
	\caption{Relative H$\beta$ and Mg\,{\it b} profiles for our ETG sample. We find declining Mg\,{\it b} profiles for most of our sample, but we find a wide range of H$\beta$ behaviours. Lines are as in \autoref{lamr}.}
	\label{lsprofiles}
	\end{center}
\end{figure}

In \autoref{mgbcorr}, we assess the well-known Mg\,{\it b} - $\sigma$ correlation out to multiple effective radii. We take an elliptical aperture of major axis radius $0.5 R_e$ along with two elliptical annuli with boundaries at 0.5$R_e$, 1$R_e$ and 2$R_e$. We take the mean of all relevant Voronoi bins as before in each case, with errors propagated accordingly. Performing least-absolute-deviation fits to our data, we find gradients $\Delta(\rm{Mg}\,{\it b}) / dex$ of 3.9, 2.6 and 4.0 respectively. We therefore find that the Mg\,{\it b} - $\sigma$ relation holds beyond the central effective radius for our sample. The changing gradients of our straight-line fits are largely driven by the galaxies NGC 2764 and NGC 3626, which both have flat Mg\,{\it b} gradients and low Mg\,{\it b} values; if we exclude these two galaxies, we obtain gradients of 4.2, 4.0 and 4.1 from the straight-line fits.

\begin{figure}
\begin{center}
	\includegraphics[trim = 2cm 5cm 0cm 12cm, scale=0.6]{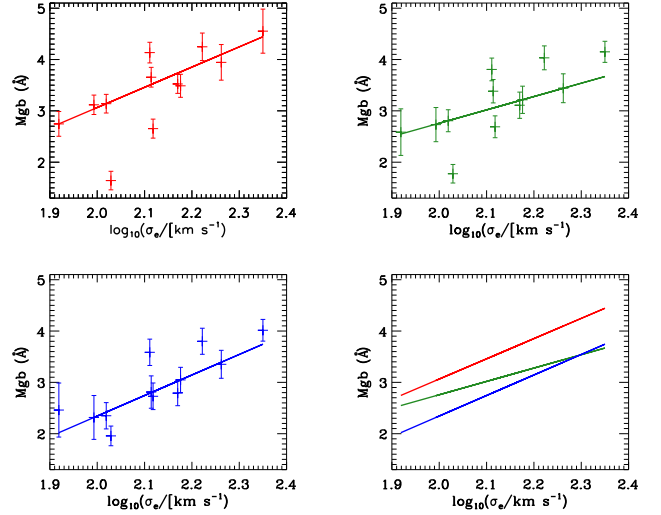}
	\caption{Mg\,{\it b}-$\sigma$ relation computed over a 0.5$R_e$ aperture (red; top left window), a 0.5-1$R_e$ annulus (green; top right window) and a 1-2$R_e$ annulus (blue; bottom left window). The thick lines show a least-absolute-deviation straight line fit for each set of points. The bottom right window shows the three straight line fits together. Values of $\sigma_e$, the velocity dispersion calculated over a 1$R_e$ aperture, are taken from \citet{cappellari2013} and references therein. We find that the Mg\,{\it b}-$\sigma$ relation holds over multiple effective radii in our sample.}
	\label{mgbcorr}
	\end{center}
\end{figure}

\section{Stellar population modelling}\label{popsppxf}

In this section, we use our emission-cleaned spectra to calculate ages, metallicities [Z/H] and abundance ratios for our sample of galaxies, with NGC 2764 excluded as discussed previously. We perform full spectral fitting of the gas-cleaned spectra using pPXF, thereby making full use of the information offered by the spectra.

We de-redshifted the galaxies' spectra and then combined them into a series of central apertures and elliptical annuli, in order to ensure sufficiently high S/N when performing the fit. We used ellipticities equal to the global ellipticity for each galaxy, with aperture/annulus boundaries at 0.5$R_e$, 1$R_e$, 1.5$R_e$ and 2.5$R_e$; for each aperture or annulus, we then combine all Voronoi-binned spectra for which the luminosity-weighted centre falls within the annulus. We discounted the outer annulus for galaxies with $R_{\rm max} < 2.5$. We then cleaned the combined spectra of  \textit{all} detected gas emission rather than using the A/N limits discussed previously. This is because the detected emission often went below the conventional A/N limits at the edges of our FOV, resulting in significant under-subtraction of the emission (and so notable residuals in the pPXF fits) when the spectra were combined. 

We performed spectral fitting on the combined spectra by using pPXF to fit linear combinations of Simple Stellar Population (SSP) models from the MILES SSP library \citep{vazdekis2010}. We used the solar-scaled ($[\alpha / Fe] = 0$) and alpha-enhanced ($[\alpha / Fe] = 0.4$)  MILES models presented in \citet{vazdekis2015}, spaced approximately logarithmically in age and metallicity. We did not use exact log-spacing for age and metallicity, due to the models themselves not being available with such spacing. We used model ages of 0.5 Gyr, 0.7 Gyr, 1 Gyr, 1.5 Gyr, 2.25 Gyr, 3.25 Gyr, 4.5 Gyr, 6.5 Gyr, 9.5 Gyr and 14Gyr; we used model metallicity $[Z/H]$ values meanwhile of -2.27, -1.79, -1.49, -1.26, -0.96, -0.66, -0.35, 0.06 and 0.4. We also tracked the stellar mass $M_*$ and luminosity $L_V$ for each model, where $M_*$ includes the mass of stellar remnants but excludes the gas lost during stellar evolution, in order to derive a mass-to-light ratio ($M_*/L_V$) for each combined spectrum. 

We allowed for four kinematic moments in the fit, as well as a 10th-degree multiplicative polynomial. We used a pPXF penalty parameter of 0.2, as before. For certain galaxies, we noted sharp sky features in the combined spectra that pPXF had not previously been able to subtract out; we created narrow masks over these features in such cases. We did not broaden the spectra to match the MILES spectral resolution in this case, in order to avoid degrading our spectra unnecessarily: our combined Mitchell spectra will be broader than the MILES SSP templates already due to the stellar velocity dispersion, while the SSP templates are broadened as appropriate by pPXF as part of the pPXF fit.

The pPXF SSP fits enable us to detect multiple stellar population components, which in turn allows us to measure the star-formation history (SFH) for each galaxy. Inferring the SFH from spectra is an intrinsically degenerate process, however, as multiple combinations of stellar population components can produce near-identical observations. We account for this degeneracy by imposing a regularisation constraint, using the ``regul" keyword in pPXF, in order to force the pPXF solution towards the maximum smoothness allowed by the data \citep[see Section 3.5 of][for details.]{cappellari2017}.

 The amount of regularisation is controlled by a single regularisation parameter. We optimise this parameter for each combined spectrum in turn as follows. We first perform a pPXF fit with no regularisation and scale the errors on our spectra such that $\chi^2 = N$, where N is the number of good pixels across the spectrum.  We then choose the regularisation parameter such that $\Delta\chi^2 \simeq \sqrt{2N}$, where $\Delta\chi^2$ indicates the difference in $\chi^2$  values for the regularised and non-regularised fits. We derived errors on our stellar population values by performing 100 monte-carlo re-simulations with added Gaussian noise, using zero regularisation and zero penalty. 

We calculated light-weighted and mass-weighted values for the age, metallicity and abundance ratio of each fitted model. The light- and mass-weighted value of a given parameter is then given by:

\begin{equation}\label{popeq1}
X_{mass} = \frac{\sum_{k=1}^K w_k\times x_k}{\sum_{k=1}^K w_k}
\end{equation}  

\begin{equation}\label{popeq2}
X_{light} = \frac{\sum_{k=1}^K w_k\times x_k \times F_k}{\sum_{k=1}^K w_k \times F_k}
\end{equation}  

\noindent where $x$ represents the parameter value of a given template, $X$ the final weighted value of that parameter, $w_k$ the template weights and $F_k$ the mean flux of each template across the Mitchell wavelength range. We use logarithms of the model ages when applying these equations, due to the logarithmic spacing employed for the SSP models. We calculated $M_*/L_V$ using 

\begin{equation}\label{popeq3}
M_*/L_V = \frac{\sum_{k=1}^K w_k\times M_{k,*}}{\sum_{k=1}^K w_k \times L_{k,V}}
\end{equation}  

We compared the light-weighted stellar population parameters inferred from regularised and non-regularised pPXF fits, in order to verify that the parameters do not significantly depend on the regularisation. We show the results of this comparison in \autoref{regnoreg}. We find near one-to-one agreement overall between the two sets of values, though with a slight trend towards lower ages and higher metallicities. The fits to the outermost binned spectrum of UGC 03960 produce a significant anomaly in terms of age, with the non-regularised fit producing an age of 14Gyr and the regularised fit yielding an age of 5.4Gyr. This is caused by the non-regularised fit being at the edge of our model grid's parameter space, along with the relatively large degree of regularisation applied to the regularised fit: the regularisation increases the relative weights of young SSP models, thus lowering the light-weighted age.

\begin{figure}
\begin{center}
	\includegraphics[trim = 3cm 2cm 0cm 12cm, scale=0.5]{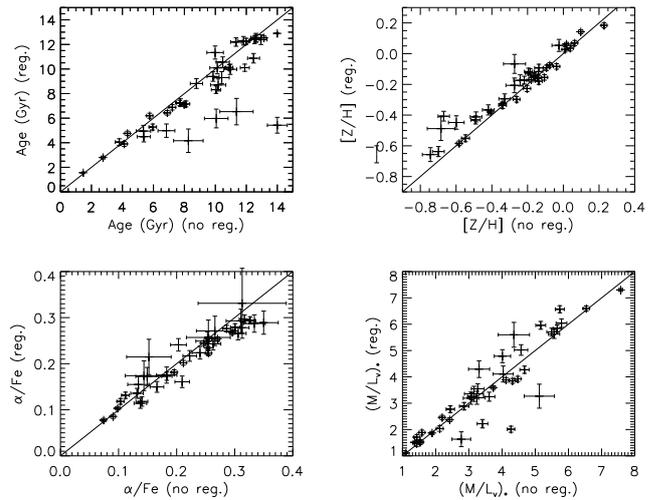}
	\caption{Comparison between stellar population parameters obtained from regularised and non-regularised pPXF fits, with black solid lines indicating one-to-one relations. We find good agreement between the two sets of fits, showing that our regularisation scheme does not significantly affect the derived values.}
	\label{regnoreg}
	\end{center}
\end{figure}

In \autoref{atlasmitssp}, we compare light-weighted and mass-weighted ages calculated for $1R_e$ apertures with the ATLAS\textsuperscript{3D} values reported in \citet{mcdermid2015}, along with the $[Z/H]$ values inferred from our respective studies. The mass-weighted values in \citet{mcdermid2015} were calculated using a very similar procedure to ours, though with MIUSCAT SSP models \citep{vazdekis2012} rather than MILES models.  We compare our light-weighted values with values calculated from fits to absorption line indices, which used \citet{schiavon2007} models. Our ages compare well with \citet{mcdermid2015} and are consistent with a one-to-one relation; however, our light-weighted $[Z/H]$ values are higher by an average of 0.13, while our mass-weighted $[Z/H]$ is higher by an average of 0.32.

\begin{figure}
\begin{center}
	\includegraphics[trim = 3cm 2cm 0cm 12cm, scale=0.5]{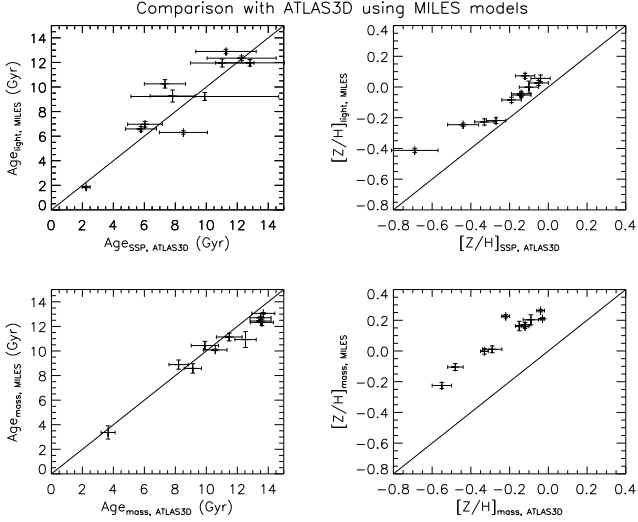}
	\caption{Comparison between stellar population parameters calculated around $1R_e$ apertures with those reported in \citet{mcdermid2015} from ATLAS\textsuperscript{3D} data. The black solid lines indicate one-to-one relations. We find good agreement in terms of age and light-weighted $[Fe/H]$, but find significant offsets in terms of mass-weighted $[Fe/H]$.}
	\label{atlasmitssp}
	\end{center}
\end{figure}

To test the effect of our choice of model library, we also performed pPXF fits to $1R_e$ apertures using SSP models from the MIUSCAT library. As in \citet{mcdermid2015}, we use a MIUSCAT model grid spanning a regular grid of log(age) and metalicity, using ages of 0.1-14 Gyr and metalicities $[Z/H]$ of -1.71 to 0.22, giving 264 models in total. We optimise the amount of regularisation in the same manner as described previously. We compare the results of these fits to the \citet{mcdermid2015} values in \autoref{atlasmitssp2}. We find near one-to-one agreement in the galaxies' ages as well as in their light-weighted metallicity, though we obtain mass-weighted $[Z/H]$ values that are higher on average by 0.08 with respect to ATLAS\textsuperscript{3D}. As such, the metallicity offsets seen in \autoref{atlasmitssp} appear to be largely due to our choice of SSP models.

\begin{figure}
\begin{center}
	\includegraphics[trim = 3cm 2cm 0cm 12cm, scale=0.5]{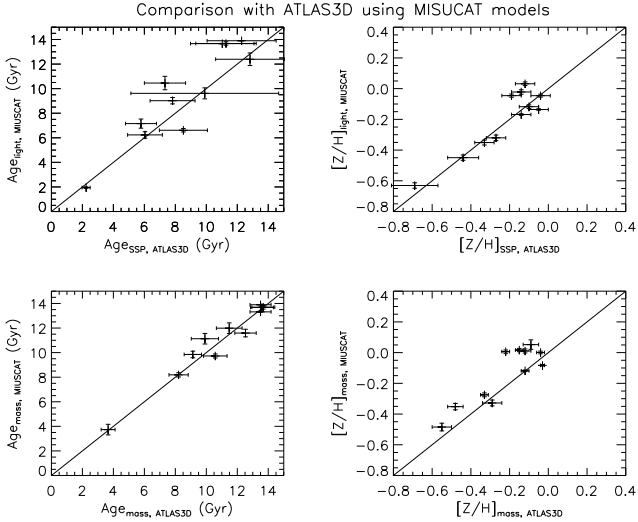}
	\caption{Comparison between stellar population parameters calculated around $1R_e$ apertures with MIUSCAT models with those reported in \citet{mcdermid2015} from ATLAS\textsuperscript{3D} data. The black solid lines indicate one-to-one relations. We find good agreement in terms of age and light-weighted $[Z/H]$, but continue to observe a mild offset in mass-weighted $[Z/H]$.}
	\label{atlasmitssp2}
	\end{center}
\end{figure}

We present profiles of light-weighted age and metallicity and $[\alpha/Fe]$ ratio in \autoref{agemetallight}, in which the radius is given as the flux-weighted average radius for a given combined spectrum. We find negative metallicity gradients for all tested galaxies. We find age gradients that are negative on average, but detect a wide degree of variation between individual systems. NGC 3626 is particularly notable, both due to its low central age and its strong positive age gradient. We note here that the age resolution of our model grid is very coarse at 9 Gyr and above, and so caution against over-interpreting the age results in that region.  We also find our galaxies to have slightly positive $[\alpha/Fe]$ gradients overall, though we caution that our grid sampling in terms of $[\alpha/Fe]$ is likewise coarse.

\begin{figure}
\begin{center}
	\includegraphics[trim = 1cm 2cm 0cm 9cm, scale=0.6]{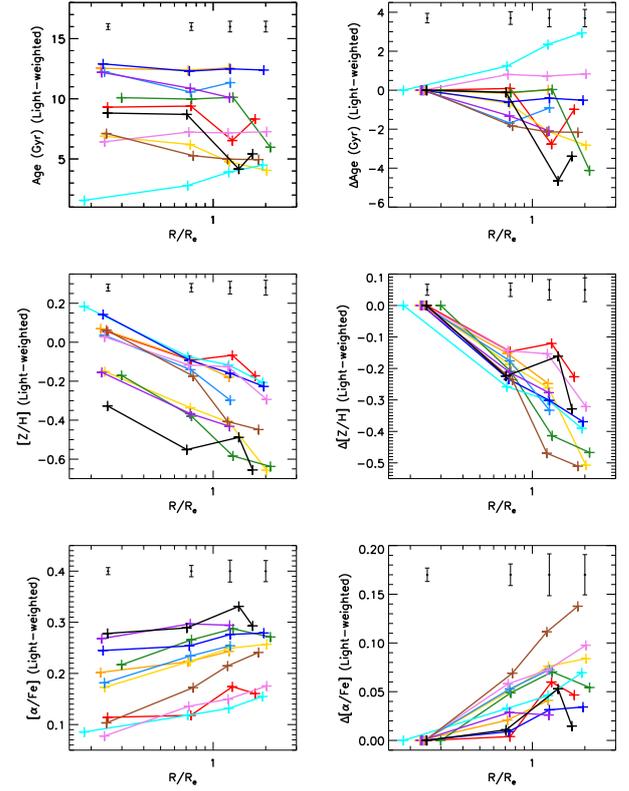}
	\caption{Absolute (left) and relative (right) profiles of light-weighted age (top), metallicity (middle) and $[\alpha/Fe]$ abundance ratio, derived from pPXF fits to combined spectra. Lines are as in \autoref{lamr}. The black error bars show the mean error at a given aperture or annulus position.}
	\label{agemetallight}
	\end{center}
\end{figure}

We present profiles of $M_*/L_V$ \autoref{mllight}.  We find three broad behaviours in the $M_*/L_V$ profiles: we find a positive gradient for NGC 3626, a near-flat gradient for NGC 5631, and negative gradients for the remainder of the sample. This is interesting to consider in the context of ETG dynamics, given that $M_*/L$ is typically assumed to be constant in dynamical modelling studies; we explore this point further in \autoref{disc}.  \citet{tortora2011} have previously investigated $M_*/L$ variations by modelling SDSS imaging data out to 1$R_e$, and report age gradients within galaxies to be the key factor in $M_*/L$ behaviour. We tested for a similar relationship here. We first calculated global parameter gradients for each galaxy by performing least-absolute-deviation fits to our profiles, in log space. We then compared the gradients in light-weighted age, metallicity and abundance ratio with the gradients derived for $M_*/L_V$; we show the results of this in \autoref{mlgrads}, in which we also show the linear Pearson correlation coefficient R in each instance. We find a strong correlation between the gradients for $M_*/L_V$ and age, with only weak correlations between the gradients in $M_*/L_V$  and those for metallicity or abundance ratio; as such, the changes in age appear to be the main cause of the $M_*/L_V$ gradients. We find median gradients of -0.047 dex/dex, -0.38 dex/dex and 0.056 dex/dex for galaxies' light-weighted age, metallicity and abundance ratio respectively. We also find a median $M_*/L_V$ gradient of -0.16 dex/dex.

\begin{figure*}
\begin{center}
	\includegraphics[trim = 2cm 6cm 0cm 15cm, scale=0.9]{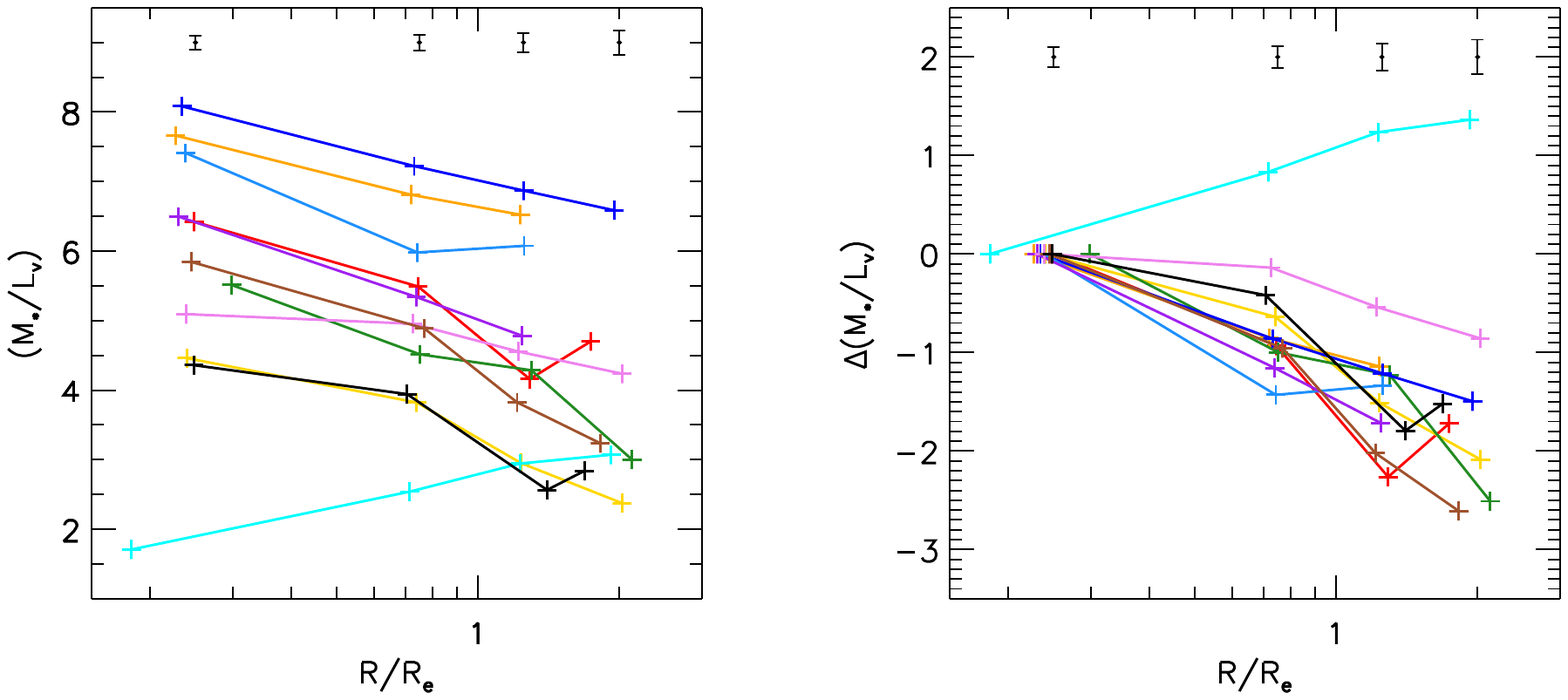}
	\caption{Absolute (left) and relative (right) profiles of stellar mass-to-light ratio $M_*/L_V$ , derived from pPXF fits to combined spectra. Lines are as in \autoref{lamr}, with error bars showing the mean error at a given aperture or annulus.}
	\label{mllight}
	\end{center}
\end{figure*}

\begin{figure*}
\begin{center}
	\includegraphics[trim = 1cm 1cm 0cm 20cm, scale=0.9]{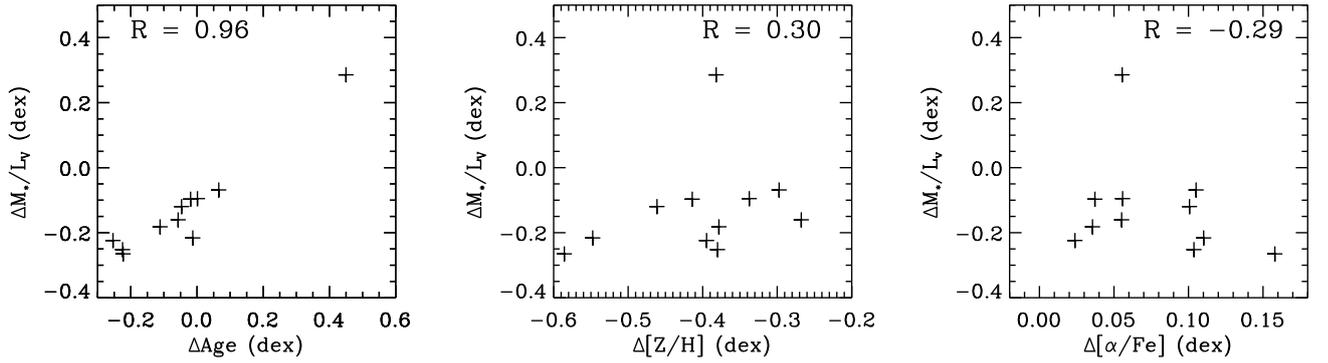}
	\caption{Gradients in $M_*/L_V$ plotted as a function of age, metallicity and abundance ratio. Individual parameters are as discussed in the text. The $M_*/L_V$ gradient correlates strongly with age while showing little dependence on metallicity or abundance ratio, implying that age variations are the main driver of the $M_*/L_V$ gradients in our results.}
	\label{mlgrads}
\end{center}
\end{figure*}

In \autoref{agegrads}, we plot the difference in light-weighted and mass-weighted values obtained for the galaxies' stellar age profiles. Most of our galaxies show mild differences of approximately $1 Gyr$, with little evolution with radius; this is indicative of the galaxies containing single dominant stellar populations. NGC 6798 is an exception here, showing large differences at low radii. NGC 3626 notably does \textit{not} show differences significantly higher than the remainder of the sample, despite appearing significantly younger than NGC 6798 in a light-weighted sense.

\begin{figure}
\begin{center}
	\includegraphics[trim = 3cm 6cm 0cm 15cm, scale=0.5]{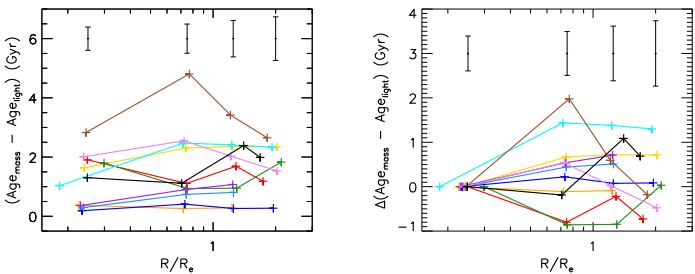}
	\caption{Profiles of the difference between light-weighted and mass-weighted galaxy ages. Line colors are as in \autoref{lamr}. We find mild differences of approximately 1-2 Gyr in most cases, but we find a larger difference for NGC 6798 (brown line).}
	\label{agegrads}
	\end{center}
\end{figure}

In \autoref{3626ssp}, we present the SSP mass-weight maps from the regularised SSP pPXF fits to NGC 3626 in terms of age and metallicity. We find the fits to predict a significant young stellar population component in the centre of NGC 3626, which becomes increasingly sub-dominant at larger radii. Near the centre of NGC 3626, the young stellar populations are significant in both a mass-weighted and light-weighted sense. Further out, such populations contribute little to the mass but continue to contribute to light-weighted parameters due to the relative brightness of young stars; which explains the behaviour for this galaxy seen in \autoref{agegrads}, in which the difference between light-weighted and mass-weighted age increases with radius. 

\begin{figure}
\begin{center}
	\includegraphics[trim = 2cm 2cm 0cm 14cm, scale=0.5]{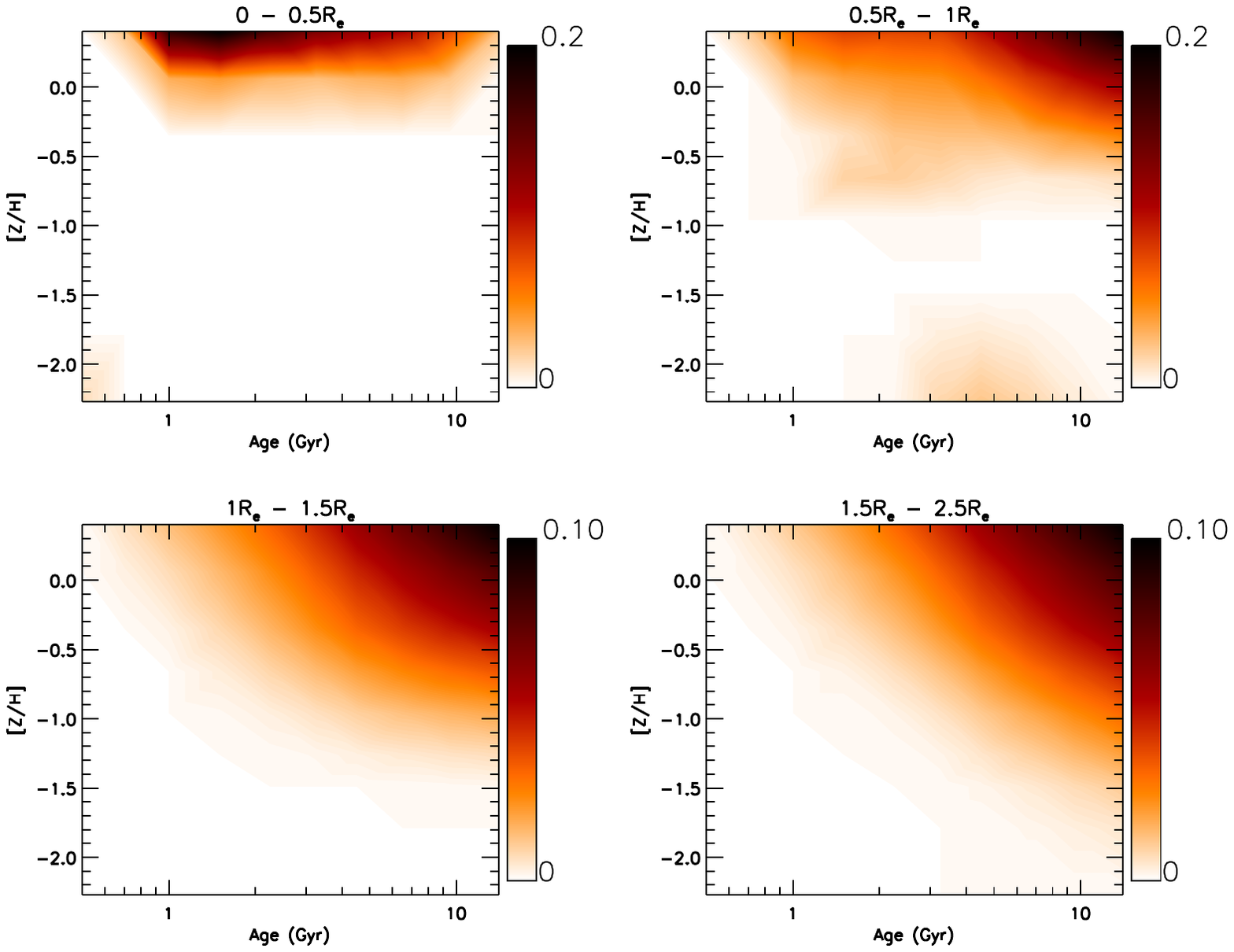}
	\caption{Mass-weight maps from pPXF SSP model fits to NGC 3626. At the centre, this galaxy contains both significant young and old populations (top left window); at larger radii, the fits are consistent with the galaxy being dominated by old stars.}
	\label{3626ssp}
	\end{center}
\end{figure}

We present NGC 6798's mass-weight maps in \autoref{6798ssp}, in which we find NGC 6798 to be dominated by an old population throughout the tested FOV. A young sub-population is also present, however, which is emphasised by the spike in this galaxy's profile seen in \autoref{agegrads}; this sub-population becomes significantly more apparent when flux-weighting is considered, as shown in \autoref{6798ssp2}.

\begin{figure}
\begin{center}
	\includegraphics[trim = 2cm 2cm 0cm 14cm, scale=0.5]{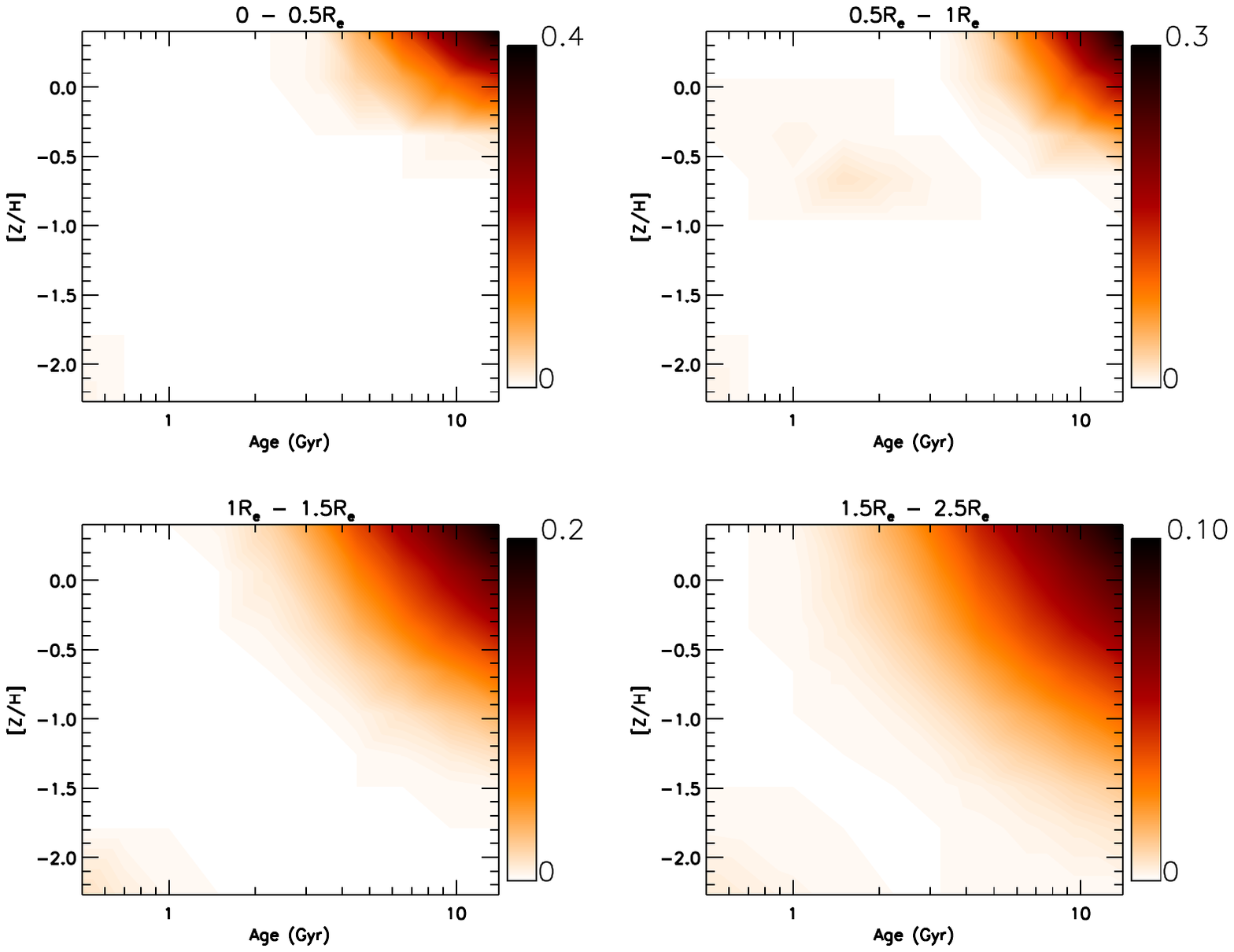}
	\caption{Mass-weight maps from pPXF SSP model fits to NGC 6798. We find this galaxy's mass to be dominated by old stars across the tested FOV, with signs also of younger sub-populations as seen in the right-hand windows.}
	\label{6798ssp}
	\end{center}
\end{figure}

\begin{figure}
\begin{center}
	\includegraphics[trim = 2cm 2cm 0cm 14cm, scale=0.5]{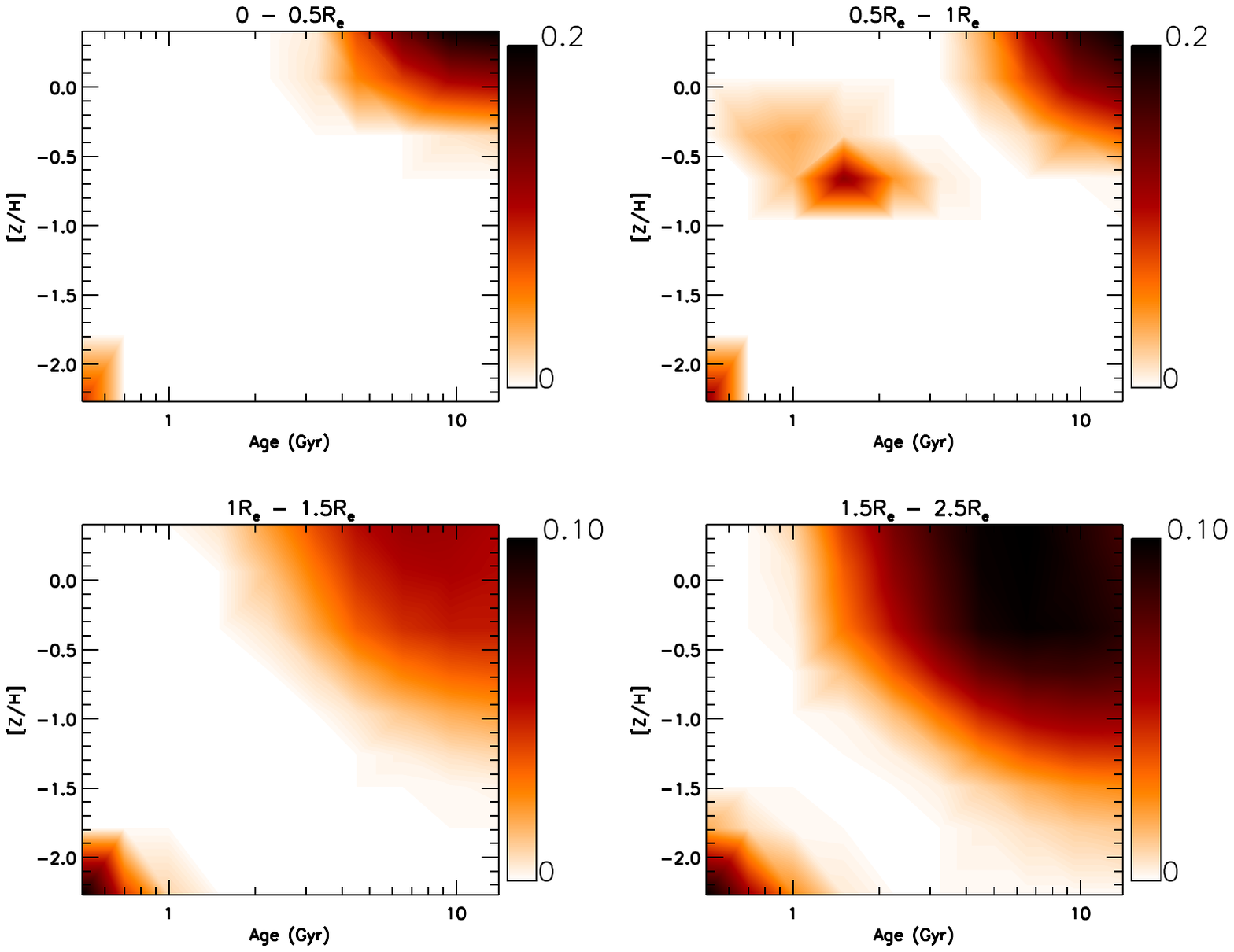}
	\caption{Flux-weight maps from pPXF SSP model fits to NGC 6798. A young sub-populations of stars is apparent in the fits.}
	\label{6798ssp2}
	\end{center}
\end{figure}

\section{Discussion}\label{disc}

In the previous sections, we presented stellar kinematics for our sample of ETGs. We probed for any stellar kinematics transitions beyond $1R_e$, using the $\lambda_R$ parameter as a proxy for angular momentum. We extracted ionised gas fluxes and kinematics, and showed a significant fraction of our galaxies to contain ionised gas that is misaligned with the galaxies' stars. Lastly, we performed spectral fitting with SSP models in order to investigate the galaxies' stellar populations.

Visually, we see no dramatic changes in the stellar kinematics beyond $1R_e$: galaxies with significant rotation at $1R_e$ remain fast-rotating as far out as we observe, and slow-rotating galaxies likewise remain slow-rotating. Our $\lambda_R$ profiles support this view, with no dramatic changes seen beyond $1R_e$. Crucially, we observe no \textit{decreases} of these parameters with position beyond the central half-light radius. We therefore confirm that the fast/slow rotator classification \citep{emsellem2007} seems to generally characterise well the overall dynamical status of our sample galaxies' stellar populations. Our kinematics results are different to those of \citet{arnold2014}, who report declines in specific angular momentum beyond 1$R_e$ in a few cases and who argue this to signify a transition to stellar halo dominated outskirts and so late dry accretion, but are qualitatively similar to the those of \citet{raskutti2014}.

The apparent contrast between our results and those of \citet{arnold2014} is due to differences between our sample and theirs. 8 out of our 9 FRs would be classified as lenticular galaxies on the basis of their Hubble T-type parameter \citep[T-type $>$ -3.5;][]{paturel2003}; the majority of  \citet{arnold2014} FRs would be classified as ellipticals on this basis and include multiple disky ellipticals, or E(d)s. E(d)s are characterized by a stellar disk which only dominates the light within the central regions, leading to drops in observed stellar rotation beyond $\sim 1 R_e$. Most FRs are instead dominated by rotation out to large radii, meaning that drops in momentum are not generally observed \citep[e.g.][]{coccato2009}. This difference in the kinematic behaviour of fast rotators at large radii is illustrated in Figure 3 of \citet{cappellari2016}, and our observations appear consistent with this general picture.

\citet{rottgers2014} have found in their simulations that in-situ and ex-situ formed stars follow similar orbit distributions; as such, it is difficult to rule out late dry accretion from the kinematics alone. However, we note that the $\lambda_R$ profiles for our FRs are very similar in shape to those presented in \citet{naab2014}, which are based on the same simulations as \citet{rottgers2014}, for galaxies that have experienced late ($z < 2$) gas-rich mergers and/or gradual dissipation. As such, we find our $\lambda_R$ profiles to be consistent with gas-rich histories for these galaxies. We reiterate at this point that we selected our sample based on the presence of \HI ; as such, the similarity of our profiles to those of the gas-rich \citet{naab2014} galaxies is unsurprising but is nonetheless encouraging.

The ionised gas content of these galaxies also supports the idea of several of them them having experienced gas-rich events in their pasts, with significant misalignments between the gaseous and stellar components in several cases. We find counter-rotating gas in NGC 3626, NGC 5631 and NGC 6798. The gas kinematics for NGC 2685 and NGC 3998 both show signs of warps, with NGC 3998's case in particular resembling the simulated gas disk in \citet{vandevoort2015}, while NGC 3522 features a polar-rotating gas disk. Such features imply that the gas in these galaxies was accreted after the galaxies themselves were formed, or else was disturbed and then subsequently re-accreted. Such arguments could also hold for some co-rotating ionised gas structures, since an accreted gas disk should ultimately align with its host galaxy's stars in half of all cases \citep[e.g.][]{katkov2015}; the time-scales of such alignments (or anti-alignments) will vary from object to object, but could potentially be several Gyr in the case of extended accretion \citep{vandevoort2015}.

Our stellar population modelling results also appear to discount late dry major mergers for the eleven galaxies that we modelled, as well as late major mergers in general. Here, we find that the metallicity continues to decline well beyond the central effective radius, with little change seen in metallicity gradients beyond 1$R_e$. Gas-rich major mergers are predicted to steepen metallicity gradients within approximately 1$R_e$ \citep{hopkins2009a}, which is not supported by our modelling. Gas-poor major mergers, meanwhile, are predicted to flatten metallicity gradients; \citet{kobayashi2004}, for instance, find a mean typical metallicity gradient of -0.19 dex/dex from their major merger simulations. Our stellar population modelling yields metallicity gradients that are typically steeper than predicted to result from dry major mergers, with our stellar kinematics likewise inconsistent with late dry major merging. Our results are therefore consistent with our galaxies having not experienced major mergers at late times. 

Our metallicity gradients also agree well with those from the cosmological simulations of \citet{hirschmann2015} in the case when galactic winds are included, which yield metallicity gradients of -0.35 dex/dex; their wind-free simulations, by contrast, yield a mean gradient of just -0.11 dex/dex. \citet{hirschmann2015} find that galactic winds lower the rate of stellar accretion in their models while also lowering the metallicity of accreted stars, with stellar accretion then serving the steepen the metallicity gradient. As such, our results are not  inconsistent with dry minor mergers and/or accretion events having occurred.

Our investigation of the Mg\,{\it b}-$\sigma$ relation beyond $1R_e$ shows this relation to persist beyond the central effective radius, meanwhile, with similar Mg\,{\it b} gradients for most of our galaxies leading to correspondingly similar gradients in the Mg\,{\it b}-$\sigma$ relation out to large radii. We infer from this that the galaxies in our sample have had broadly similar evolution histories, which we ascribe in turn to our selection criteria.

We found positive $[\alpha / Fe]$ gradients for all galaxies for which we performed stellar population modelling. $[\alpha / Fe]$ has previously been shown to be linked to the formation time-scale of stellar populations, with populations with greater $\alpha$-enhancement having shorter time-scales on average \citep[e.g.][]{thomas2005,mcdermid2015}. As such, our results here indicate that the stars in our ETGs have shorter formation time-scales at larger radii. This result is similar to those that have previously been reported in bulges of galaxies across the Hubble sequence \citep{jablonka2007}, as well as to the vertical gradients reported for barred disk galaxies \citep{molaeinezhad2017}.

We found negative age gradients in most cases, with a strong positive age gradient for NGC 3626. NGC 3626 is relatively young compared to most ETGs, being younger than 5 Gyr throughout the tested FOV in a light-weighted sense. We also compared the light-weighted stellar ages of our galaxies with the mass-weighted values, finding a significant difference in the case of NGC 6798. We demonstrated this to be due to young bright sub-populations of stars, which contribute much to a galaxy's luminosity but only a little to its overall mass budget.  Such findings could indicate that galaxies such as NGC 3626 and NGC 6798 are undergoing quenching for the first time, or else could indicate that they are experiencing a degree of rejuvenation. Given the presence of misaligned ionised gas in several of our galaxies, we argue that the latter interpretation is more likely: these galaxies experience gas-rich interaction events which subsequently triggered some amount of star formation. Even a small amount of star formation will significantly decrease the SSP light-weighted age result due to the brightness of young stars \citep[e.g.][]{scott2013}; as these galaxies age, this bright young population will fade, causing their age profiles to more closely resemble the older ETGs in our sample.

Deep imaging has also yielded signs of past interactions for several of the galaxies in our sample. \citet{duc2015} report finding signs of recent wet merging in NGC 2685 and NGC 2764; they further report tidal disturbances in NGC 680, a possible disrupted satellite 50 kpc north of NGC 3522, and multiple internal shells within NGC 5631. As such, we find our results to be consistent with the implications of the \citet{duc2015}  imaging.

Of the ten galaxies classified as ``D" in \citet{serra2012}, nine show signs of some past disturbance in their \HI\ morphologies. NGC 3626 and NGC 6798 both contain counter-rotating \HI\ discs,  while other \HI\ structures within our ``D" galaxies display warps, lopsidedness and/or kinematic misalignments with respect to galaxies' stellar kinematics. NGC 5582 is the one exception here, with its \HI\ content consisting of a large-scale ring that is well-aligned with the stellar content. The large-scale \HI\ content of these ten galaxies appears to rule out significant late-time major merging, since this would disrupt the \HI\ disc structure, but also appears to signify the need for interaction events in nearly all of these galaxies' pasts; this agrees well with the picture we have derived from our IFU data, as well as that which can be implied from deep optical imaging.

NGC 680 and NGC 1023 are both classified as ``u" in \citet{serra2012}, indicating substantially unsettled \HI\ morphology; it is therefore harder to constrain the pasts of these two systems on the basis of \HI\ alone. However, these galaxies do not display significantly different behaviour from the rest of the sample in terms of their $\lambda_R$ profiles or stellar populations, and so we argue that these galaxies most likely experienced one or more gas-rich interaction events in a similar manner to the rest of our sample.

Lastly, we found our SSP models to support negative $M_*/L_V$ gradients for the majority of our sample. Such gradients could have important implications for dynamical modelling of ETGs. Dynamical models of ETGs generally assume a flat $M_*/L$ slope. Attempts to calculate $M_*/L$ slopes in dynamical models are few and far between, but typically produce positive $M_*/L$ gradients due to the degeneracy between dark and visible mass \citep[e.g.][]{yildirim2016}. Our results imply that dynamical models of many of our ETGs would have biases in the inferred dark matter contributions were $M_*/L$ to be assumed fixed. Given that ETGs have generally been found to contain only low amounts of dark matter within their central half-light radius \citep[e.g.][]{boardman2016,cappellari2013a}, our results demonstrate the need for allowing for M*/L gradients in dynamical models, as done for instance in \citet{mitzkus2017} and \citet{poci2017}.

An important caveat is that we did not attempt to account for dust attenuation in our SSP fits; this is because the effects of attenuation are degenerate with the continuum correction over our wavelength range. Undetected dust gradients could therefore play a role in our detected stellar population gradients and in turn, our $M_*/L_V$ gradients. It would be illuminating to perform a similar stellar population analysis for IFU spectra with wider wavelength ranges, for which the effects of dust could be accounted for in a non-degenerate way.

\section{Summary and Conclusion}\label{sum}

We have presented an IFU sample of 12 nearby \HI -detected ETGs of intermediate mass, for which we obtain stellar and gaseous kinematics. We also obtain stellar population constraints out to multiple effective radii for 11 out of the 12 galaxies. Our study represents the first IFU analysis of such objects far beyond the central effective radius, allowing for valuable comparisons to simulations that could not previously have been made. Our ETGs have also been observed previously as part of the ATLAS\textsuperscript{3D} survey, making our data greatly complimentary to high spatial resolution central IFU data provided by that survey.

We find no large transitions in the stellar kinematics beyond $1R_e$: galaxies in our sample which are FRs (SRs) at $1R_e$ retain a high (low) $\lambda_R$ amplitude over the full FOV of our data relative to their respective ellipticities. We find encouraging agreement between the observed fast-rotator kinematics and those from simulations, with the spin profiles of our fast-rotators strongly resembling those of simulated galaxies with gas-rich histories. 

Our results in terms of stellar metallicity, meanwhile, appear to rule out late major merging for our galaxies. We find metallicity gradients which persist to multiple half-light radii for the eleven galaxies we modelled, with no clear flattening beyond $1R_e$. Were these galaxies to have experienced gas-rich major mergers at late times, then we would not expect such trends to survive. Our derived metallicity gradients are typically steeper than would be expected for remnants of dry major mergers, meanwhile. Our results here therefore demonstrate complimentary nature of stellar kinematics and stellar populations for uncovering ETG's histories.

We show that a significant fraction of our galaxies contained ionised gas which is counter-rotating or else misaligned with respect to the stars; this is not a new result for these galaxies, but highlights the fact that many of these galaxies cannot have had completely passive histories. Our stellar population modelling further suggest past interactions in certain cases, with noticeable age gradients in certain ETGs as well as multiple stellar population components arising from modelling of certain objects. As such, our ionised gas measurements and stellar population modelling both produce results consistent with those obtained for the galaxies' stellar kinematics.

We also find our stellar population modelling to imply negative gradients in $M_*/L_V$ for the majority of galaxies in our sample. This could have important implications for dynamical models of ETGs, for which the mass-to-light ratio is typically assumed constant and for which the central dark matter fractions are typically found to be low. We thus argued that it could be worthwhile to consider this in future dynamical modelling works.

In this paper, we have demonstrated both the reach of our observations as well as their power in investigating ETG's histories. We have shown that all approaches of analysing our spectra paint a consistent evolutionary picture: the galaxies in our sample were likely shaped by at least one gas-rich interaction apiece, with late dry major mergers ruled out by the data. We have thus shown that a wide FOV is crucial for investigating the formation paths of ETGs, with much information to be found beyond the central half-light radius.

\section*{Acknowledgements}

We wish to thank Josh Adams, Guillermo Blanc and Jeremy Murphy for their help with the data reduction. We thank Vivienne Wild and Marc Sarzi for fruitful discussions concerning our analysis. We thank Adam Atkinson, Phil Cigan, Carina Lagerholm, Kevin Luecke and Kristina Nyland for their help with observing and we thank the staff of McDonald Observatory for their support. We thank Anne Sansom for her useful and thorough referee report, which significantly improved this paper. NFB was supported by STFC grant ST/K502339/1 during the course of this work. NFB acknowledges support from the Max Planck Institute for Astronomy in Heidelberg, Germany. AW acknowledges support of a Leverhulme Trust Early Career Fellowship. MC acknowledges support from a Royal Society
University Research Fellowship. J.~F-B. acknowledges support from grant AYA2016-77237-C3-1-P from the Spanish Ministry of Economy and Competitiveness (MINECO).

\bibliographystyle{apj}
\bibliography{Citations}

\end{document}